\documentclass[aps,pre,floats,floatfix,twocolumn,showpacs,superscriptaddress]{revtex4}

\usepackage{latexsym,amsmath}
\usepackage{amsbsy}
\usepackage[pdftex]{graphicx}
\usepackage{amssymb}
\usepackage{epstopdf}
\usepackage[usenames]{color}

\begin{document}

\title{Double percolation phase transition in clustered complex networks}
\author{Pol Colomer-de-Sim\'on}
\author{Mari\'an Bogu\~n\'a}
\affiliation{Departament de F\'isica Fonamental, Universitat de Barcelona, Mart\'{\i} i Franqu\`es 1, 08028 Barcelona, Spain}

\date{\today}
            
\begin{abstract}
The internal organization of complex networks often has striking consequences on either their response to external perturbations or on their dynamical properties. In addition to small-world and scale-free properties, clustering is the most common topological characteristic observed in many real networked systems. In this paper, we report an extensive numerical study on the effects of clustering on the structural properties of complex networks. Strong clustering in heterogeneous networks induces the emergence of a core-periphery organization that has a critical effect on the percolation properties of the networks. We observe a novel double phase transition with an intermediate phase in which only the core of the network is percolated and a final phase in which the periphery percolates regardless of the core. This result implies breaking of the same symmetry at two different values of the control parameter, in stark contrast to the modern theory of continuous phase transitions. Inspired by this core-periphery organization, we introduce a simple model that allows us to analytically prove that such an anomalous phase transition is in fact possible. 
\end{abstract}

\maketitle

\section{Introduction}
The essence of complex systems lies in the interactions among their constituents. In many cases, these interactions are organized into complex topological architectures that have a determinant role in the behavior and functionality of this class of systems. In regular lattices, dimensionality appears to be one of the most distinctive features; however, randomness and heterogeneity in the interactions of complex networked systems induce phenomena that are very different from, or that are not even observed in, regular lattices. Examples range from the absence of epidemic thresholds that separate healthy and endemic phases~\cite{Pastor-Satorras:2001ly,Lloyd:2001mz,Boguna:2003zr,Berger:2005fk,Chatterjee:2009uq,Boguna:2013kx} to the anomalous behavior of Ising-like dynamics~\cite{Bianconi:2002fj,Goltsev:2003yq,Hinczewski:2006vn,Dorogovtsev:2008kx} and percolation properties~\cite{Cohen2000,Callaway2000,Cohen2002,Newman2002,Newman2003,Vazquez2003}. 

Percolation theory has played a prominent role in understanding the anomalous behaviors observed in complex networks and, in most cases, is the common underlying principle behind these behaviors. Interestingly, the interplay between a complex network topology and different percolation mechanisms leads to phenomena that have not previously been observed in statistical physics, including a lack of percolation thresholds in scale-free networks with a degree distribution of the form $P(k)\sim k^{-\gamma}$ for $\gamma < 3$~\cite{Pastor-Satorras:2001ly,Lloyd:2001mz,Boguna:2003zr,Berger:2005fk,Chatterjee:2009uq,Boguna:2013kx}, anomalous infinite-order percolation transitions in non-equilibrium growing random networks~\cite{Dorogovtsev:2001md,Callaway:2001fk}, or cascading processes in interdependent networks~\cite{Buldyrev2010a,Son:2012lq,Baxter:2012fp}.  However, these phenomena have already been observed on random graphs with given degree distributions. Random graphs of this type are locally tree-like, that is, the number of triangles, and thus the clustering coefficient, can be neglected in the thermodynamic limit. However, the strong presence of triangles is, along with the small-world effect and heterogeneity of the degree distribution, a common and distinctive topological property of many real complex networked systems. While clustering is not a necessary condition for the emergence of any of these phenomena, the effects of clustering on the percolation properties of a network are unknown. 

Percolation in clustered networks has been widely studied. However, previous reports differ concerning the position of the percolation threshold. Some studies report that clustered networks have a larger percolation threshold than do unclustered networks due to redundant edges in triangles that cannot be used to connect to the giant component~\cite{Kiss2008,Newman2009,Miller2009,Gleeson2010a}. Other studies report that strongly clustered networks are more resilient due to the existence of a core that is extremely difficult to break~\cite{Newman2003b,Gleeson2009,Serrano2006a}. In fact, as we shall demonstrate, both arguments are correct.

In this paper, we show that strong clustering induces a core-periphery organization in the network~\cite{csermely:2013} that gives rise to a new phenomenon, namely, a ``double percolation'' transition, in which the core and periphery percolate at different points. This behavior is in stark contrast to the modern theory of continuous phase transitions, which forbids the possibility of breaking the same symmetry at two different values of the control parameter. Multiple percolation transitions have recently been reported in~\cite{Nagler:2012fr,Chen:2013rt,Chen:2013ys,Bianconi:2014}. However, in each of these cases, anomalous percolation arises as a consequence of either complex percolation protocols~\cite{Nagler:2012fr,Chen:2013rt,Chen:2013ys} or the interdependence between different networks~\cite{Bianconi:2014}, and it is never associated with the same symmetry breaking. Instead, our results are obtained with the simplest percolation mechanism, bond percolation with bond occupation probability $p$, which indicates that this double percolation transition is exclusively induced by a particular organization of the network topology.

\begin{figure}[t]
\includegraphics[width=\linewidth]{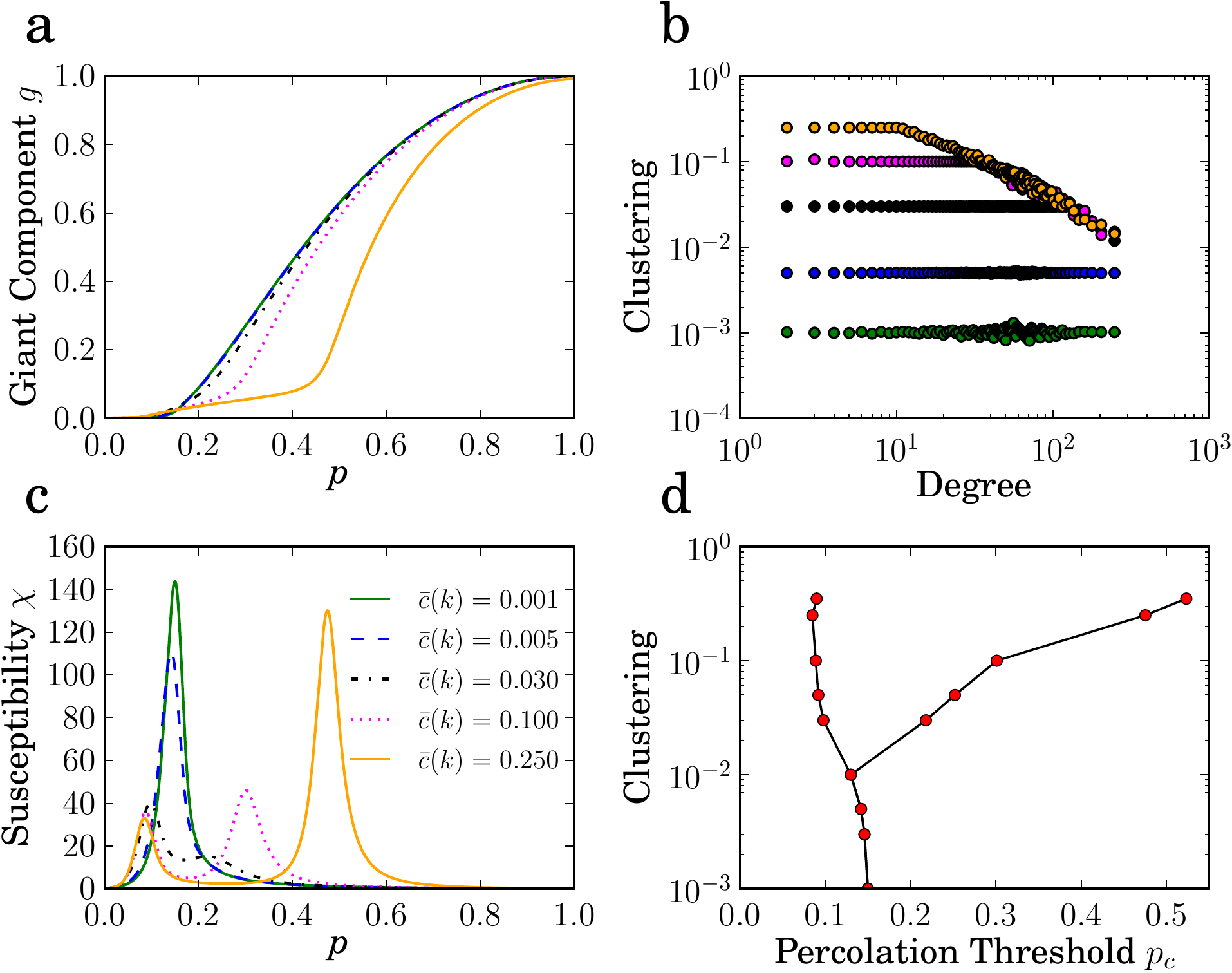}
\caption{Bond percolation simulations for networks of $N=5 \times 10^4$ nodes with a power law degree distribution, $\gamma=3.1$, and different levels of clustering. {\bf a:} Relative size of the largest connected component $g$ as a function of the bond occupation probability $p$. {\bf b:} Degree-dependent clustering coefficient $\bar{c}(k)$. {\bf c:} Susceptibility $\chi$ as a function of the bond occupation probability $p$. {\bf d:} Percolation threshold ($p_{max}$) as a function of the level of clustering.}
\label{fig:clustering}       
\end{figure}

 \section{Random graphs with a given clustering spectrum}
We can generate scale-free random graphs with a given clustering spectrum $\bar{c}(k)$ and fixed degree-degree correlations, as shown in the Appendix~\ref{appendix_A}. A preliminary analysis shows that the percolation properties depend on two network features, the heterogeneity of its degree distribution and the shape of the clustering spectrum $\bar{c}(k)$~\cite{Serrano2006a}.
For weakly heterogeneous networks  ($\gamma \gg 3$), we observe that increasing clustering in the network while keeping the degree-degree correlations fixed increases the percolation threshold and decreases the size of the giant component (see the Appendix~\ref{appendix_B}). However, the most interesting case corresponds to heterogeneous networks, typically with $\gamma<3.5$. In this work, we focus on the case of $\gamma=3.1$ and a constant clustering spectrum~\footnote{Note that this case is not the same as fixing the average clustering coefficient because, in our case, we enforce nodes of any degree to have the same local clustering. In any case, due to structural constraints, for very strong clustering it is not possible to keep $\bar{c}(k)$ constant for very large values of $k$. In this case, the algorithm generates the maximum possible clustering~\cite{Serrano:2006qj}}. This value of $\gamma$ generates scale-free heterogeneous networks but with a finite second moment, which allows us to clearly isolate the new phenomenon. The results for $\gamma \le 3$ are qualitatively similar but more involved and will be presented in a forthcoming publication.

\begin{figure}[t]
\includegraphics[width=\linewidth]{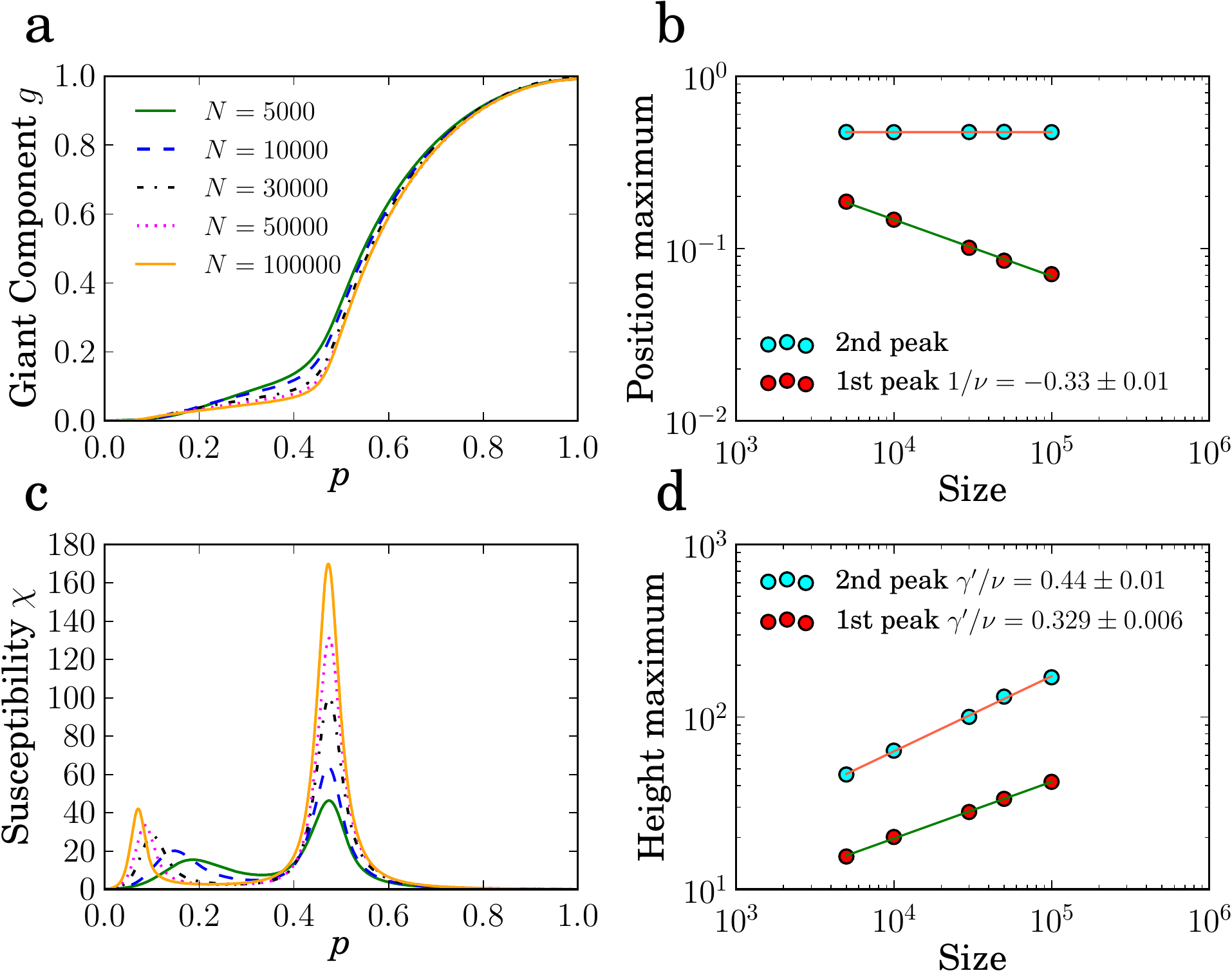}
\caption{Bond percolation simulations for networks with a power law degree distribution with $\gamma=3.1$, target clustering spectrum $\bar c(k)=0.25$, and different network sizes. {\bf a:} Relative size of the largest connected component as a function of the bond occupation probability $p$. {\bf c:} Susceptibility $\chi$ as a function of the bond occupation probability $p$. {\bf b} and {\bf d:} Position $p_{max}$ and height $\chi_{max}$ of the two peaks of $\chi$ as functions of the network size $N$. The straight lines are power-law fits, and {\bf b} and {\bf d} show the measured values of the critical exponents.}
\label{fig:FSS}
\end{figure}
\begin{figure*}[t]
\includegraphics[width=\linewidth]{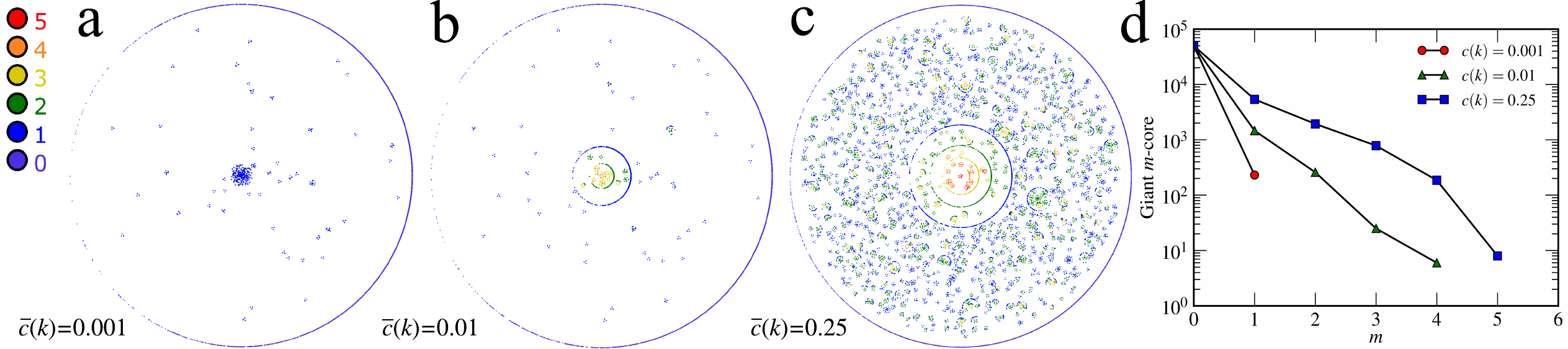} 
\caption{\small{{\bf a--c:} $m$-core decomposition of three different networks with $N=5\times 10^4$, $\gamma=3.1$, and different levels of clustering, $\bar{c}(k)=0.001,0.1,0.25$. The color code of a node represents its $m$-coreness. For instance, nodes colored violet belong to the $m0$-core but not to the $m1$-core and are said to have $m$-coreness of zero. The blue colored nodes belong to the $m1$-core but not to the $m2$-core and have $m$-coreness of 1, etc. The visual representation is as follows. The outermost circle and its contents represent the $m0$-core and therefore the entire network. If we recursively remove all edges of multiplicity $0$, we obtain the $m1$-core subgraph, which is contained within the $m0$-core. Nodes with no remaining connections do not belong to the $m1$-core, have $m$-coreness of $0$, and are located at the perimeter of the outermost circle. If the $m1$-core is fragmented into different disconnected components, they are represented as non-overlapping circles within the outermost one and with nodes of $m$-coreness of $1$ located in their perimeters (see, for instance, panels b and c). The same process is repeated for each disconnected $m1$-core, which will contain a subset of the $m2$-core, and so on. Links between nodes are not depicted for clarity. {\bf d:}  The size of the giant $m$-core as a function of $m$ for the networks shown in panels {\bf a--c}.}}
\label{fig:mcore}
\end{figure*}

Figure~\ref{fig:clustering} compares the percolation properties of networks with identical degree sequence and degree-degree correlations but with different levels of clustering. For each network, we perform bond percolation $10^4$ times using the Newman--Ziff algorithm~\cite{Newman2000} and measure the average relative size of the largest (giant) connected component, $g \equiv \langle G \rangle/N$, and its fluctuations, {\it i.e.}, the susceptibility $\chi=\left[\langle G^2 \rangle - \langle G \rangle^2\right]/\langle G \rangle$. These results are then averaged over 100 network realizations. In finite systems, a peak in the susceptibility $\chi$ indicates the presence of a continuous phase transition, and its position provides an estimate of the percolation threshold. Plots {\bf c} and {\bf d} in Fig.~\ref{fig:clustering} show new and surprising results. For low levels of clustering, there is a unique and well-defined peak in $\chi$, but increasing clustering gives rise to the emergence of a secondary peak at higher values of $p$. This result suggests the presence of a double phase transition, in which two different parts of the network percolate at different times.

To confirm this possibility, we perform finite size scaling on networks with a target clustering spectrum of $c(k)=0.25$ and different system sizes, ranging from $N=5\times 10^3$ to $N=5\times 10^5$. Plot {\bf d} in Fig.~\ref{fig:FSS} shows that the susceptibility exhibits two peaks whose maxima $\chi_{max}$ diverge as power laws, $\chi_{max}(N)\sim N^{\gamma'/\nu}$. The position of the first peak also approaches zero as a power law $p_{max}(N)\sim N^{1/\nu}$, as shown in Fig.~\ref{fig:FSS} {\bf b}, which suggests that even if the network has bounded fluctuations, $\langle k^2 \rangle < \infty$, it is always percolated in the thermodynamic limit. In contrast, the position of the second peak is nearly constant in the range of sizes we have considered. The divergence of the two peaks in the susceptibility strongly suggests that we are indeed observing two different continuous phase transitions. The first transition is between non-percolated/percolated phases, and the second transition is between two percolated phases with very different internal organizations.

\subsection{The $m$-core decomposition}
To understand the effect of clustering on the global structure of networks, we use the $m$-core decomposition developed in~\cite{Colomer-de-Simon2013}. This process is based on the concept of edge multiplicity $m$, which is defined as the number of triangles passing through an edge. We further define the $m$-core as the maximal subgraph whose edges all have at least multiplicity $m$ within it. By increasing $m$ from $0$ to $m_{max}$, we define a set of nested subgraphs that we call the $m$-core decomposition of the network.
This decomposition can be represented as a branching process that encodes the fragmentation of $m$-cores into disconnected components as $m$ is increased. The tree-like structure of this process provides information regarding the global organization of clustering in networks. To visualize this process, we use the LaNet-vi 3.0 tool developed in~\cite{Colomer-de-Simon2013} (see the caption of Fig.~\ref{fig:mcore}). Figure~\ref{fig:mcore} shows the $m$-core decomposition of three networks with $N=5\times 10^4$ nodes, the same degree sequence (with $\gamma=3.1$) and degree-degree correlations, and different levels of clustering. For low levels of clustering, the $m1$-core is very small, and thus, the $m$-core structure is almost nonexistent. As clustering increases, $m$-cores begin to develop new layers and $m_{max}$ increases. For instance, for $\bar{c}(k)=0.25$ (Fig.~\ref{fig:mcore}~{\bf c}), after the recursive removal of all links that do not participate in triangles, we obtain the $m1$-core, which is composed of a large connected cluster with a well-developed internal structure -- a core in the center of the figure -- and a large number of small disconnected components -- a periphery. This result indicates that even if the network is connected, by iteratively removing all edges with multiplicities of zero, we are left with a small but well-connected subgraph and the reminder of the network is fragmented.

\begin{figure}[t]
\includegraphics[width=\linewidth]{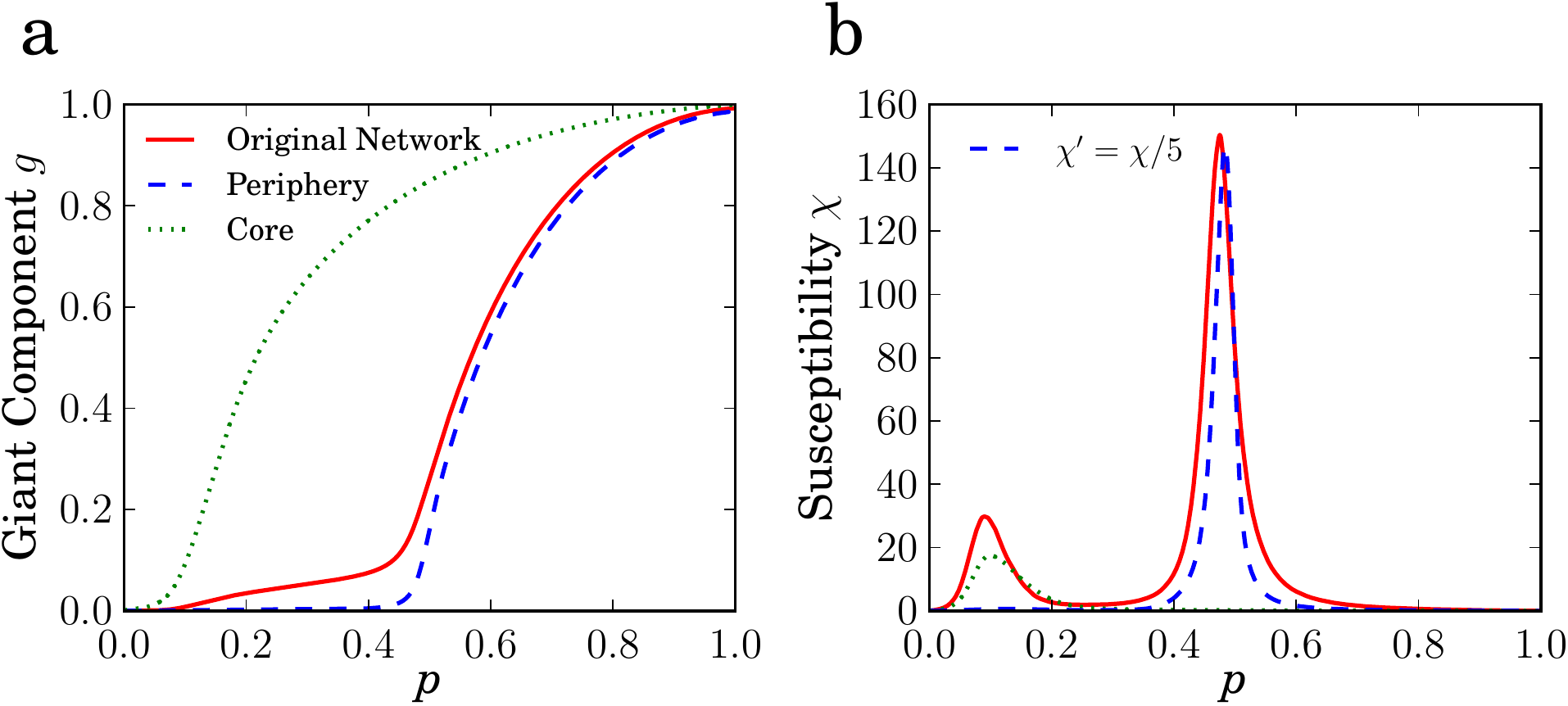} 
\caption{\small{Bond percolation simulations of the core and periphery of a network with $N=5\times 10^4$, $\gamma=3.1$, and target clustering spectrum $\bar c(k)=0.25$. The bond occupation probability to separate the core is $p=0.2$. The susceptibility curve of the periphery (dashed blue line) has been divided by $5$ for ease of comparison.}}
\label{fig:core_vs_out}
\end{figure}

The aforementioned result suggests that the two peaks in the susceptibility could be related to this core-periphery organization. Both parts would percolate at different times, first the core and then the periphery, and hence have their own percolation thresholds. To test this hypothesis, we perform bond percolation on the network with a bond occupation probability of $p$ between the two peaks. The giant component at this value of $p$ defines a subgraph that we identify with the core and that roughly corresponds to the core observed in Fig.~\ref{fig:mcore}~{\bf c} (see Appendix~\ref{appendix_C}). We then extract the latter core subgraph from the original network, and the remaining network is thus identified with the periphery.
Once the core and periphery are isolated, we perform bond percolation on both components independently and compare the results with the original network. Figure~\ref{fig:core_vs_out} shows that the core percolates precisely at the point where the first peak appears in the original network, whereas the periphery percolates at the second peak.

 \section{The core-periphery random graph: a simple model showing a double percolation transition}
The modern theory of continuous phase transitions states that, in a connected system, it is not possible to break the same symmetry at two different values of the control parameter. In our context, this statement implies that it is not possible to have two genuine percolation transitions at two different values of $p$. It is then unclear whether the second peak observed in our simulations corresponds to a real percolation transition or to a smeared transition, with the percolated core acting as an effective external field that provides connectivity among nodes in the periphery. 

Unfortunately, strongly clustered networks cannot be studied analytically. However, we can devise a system with a core-periphery organization similar to that induced by strong clustering. Let us consider two interconnected Erd\"os-R\'enyi random graphs, a core and a periphery, of average degrees of $\bar{k}_c$ and $\bar{k}_p$, respectively. The relative size is $r=N_c/N_p$, and the average numbers of connections of a node in the core to nodes in the periphery (and vice versa) are $\bar{k}_{cp}$ and $\bar{k}_{pc}=r \bar{k}_{cp}$, respectively. To model a core-periphery organization, we chose $r<1$ and $\bar{k}_c>\bar{k}_p \gg \bar{k}_{cp}$. The relative size of the giant component of the combined network is
\begin{equation}
g(p)=\frac{r}{1+r}g_c(p)+\frac{1}{1+r}g_p(p),
\label{eq:1}
\end{equation}
where $g_c(p)$ and $g_p(p)$ are the solution of the system of transcendent equations
\begin{equation}
\left.
\begin{array}{rcl}
g_c(p)&=&1-\displaystyle{e^{-p\bar{k}_c g_c(p)-p\bar{k}_{cp} g_{cp}(p)}}\\
g_{cp}(p)&=&1-\displaystyle{e^{-p\bar{k}_{pc} g_{pc}(p)-p\bar{k}_{p} g_{p}(p)}}\\
g_{pc}(p)&=&1-\displaystyle{e^{-p\bar{k}_{cp} g_{cp}(p)-p\bar{k}_{c} g_{c}(p)}}\\
g_p(p)&=&1-\displaystyle{e^{-p\bar{k}_p g_p(p)-p\bar{k}_{pc} g_{pc}(p)}}.
\label{eq:2}
\end{array}
\right\}
\end{equation}
The derivation of these equations can be found in the Appendix~\ref{appendix_D}. From here, it readily follows that $g_c$ and $g_p$ must be either both different from zero or equal to zero, implying that there is generally only one percolation transition, whereas at $p \approx \bar{k}_p^{-1}$, there is a crossover effect due to growth of the periphery.

This result is true if the coupling between the core and periphery is macroscopic, that is, the number of connections between the two structures is proportional to the size of the system such that $\bar{k}_{cp}$ and $\bar{k}_{pc}$ are constants in the thermodynamic limit. Instead, suppose that the number of connections among nodes in the core and periphery scales sub-linearly with the system size, {\it i.~e.}, as $N^{\alpha}$ with $0<\alpha<1$. In this case, $\bar{k}_{cp}$ and $\bar{k}_{pc}$ are zero in the thermodynamic limit: thus, $g_c$ and $g_p$ become decoupled in Eq.~(\ref{eq:2}) such that $g_c$ can be different from zero while $g_p=0$. However, when both the core and periphery have a giant connected component as isolated networks, the combined network forms a single connected component because there is an infinite number of connections between each part.

The effect of such structure on bond percolation is as follows. When the bond occupation probability is increased from $p=0$, the first phase transition occurs at $p=\bar{k}_c^{-1}$, where the core percolates. In the range $\bar{k}_c^{-1}<p<\bar{k}_p^{-1}$, the number of nodes in the periphery connected through the giant component of the core scales as $N^{\alpha}$; therefore, its fraction vanishes in the limit $N \gg 1$. Once we reach $p=\bar{k}_p^{-1}$, a percolating cluster is formed in the periphery that becomes macroscopic as we increase $p$ by an infinitesimal amount. At this moment, and not before, the giant clusters in the periphery and core become connected.
Thus, we have a double percolation transition defined by a regular transition at $p=\bar{k}_c^{-1}$ and the sudden emergence at $p=\bar{k}_p^{-1}$ of a macroscopic subgraph in the periphery with two types of connectivity; namely, each pair of nodes in this subgraph can be connected not only by a path going through the core but also by a path composed exclusively of nodes outside the core.

\begin{figure}[t]
\includegraphics[width=\linewidth]{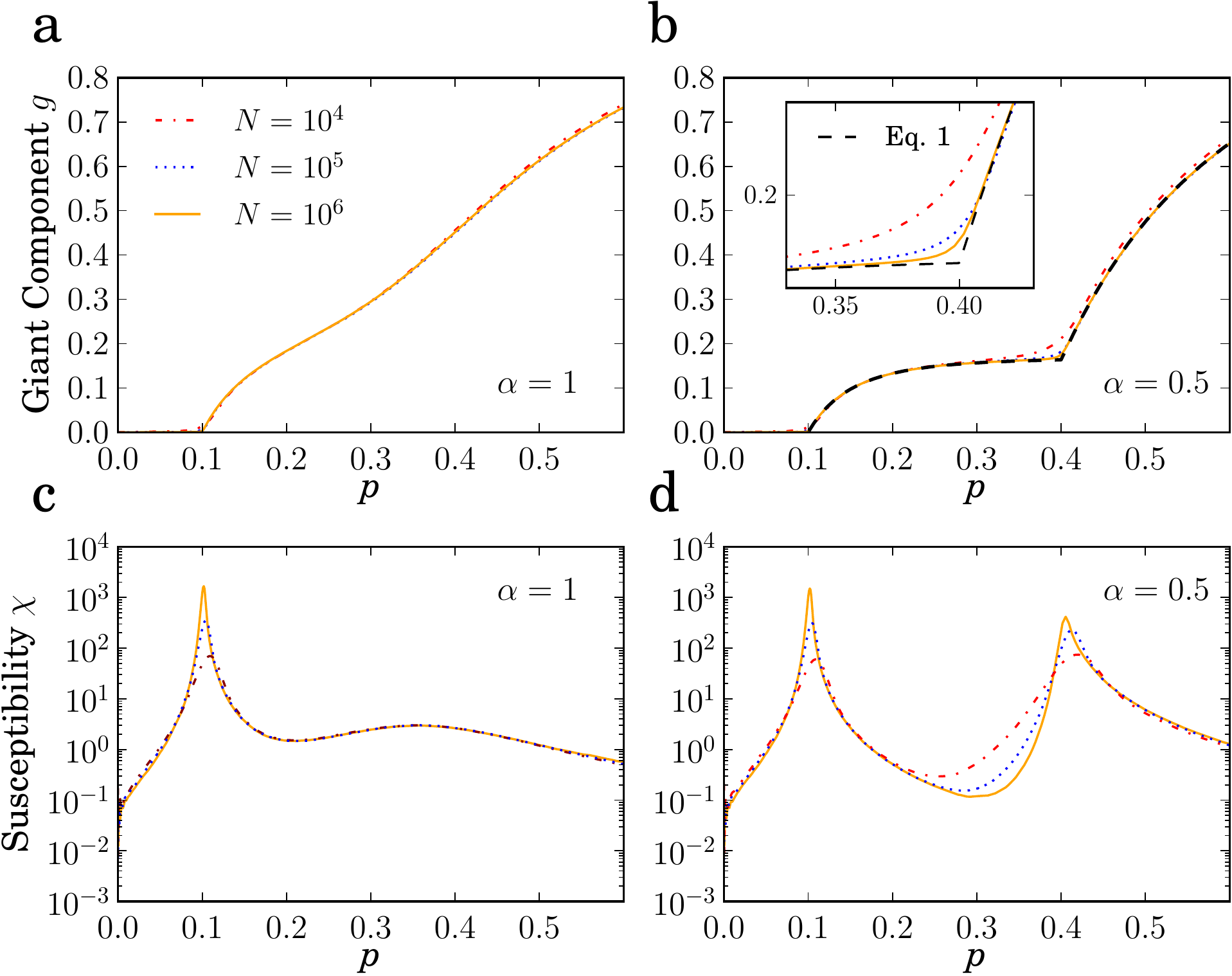} 
\caption{\small{Bond percolation simulations for the core-periphery random graph model with $\alpha=1$ (left column) and $\alpha=0.5$ (right column). In both cases, the core has an average degree of $\bar{k}_c=10$ and the periphery $\bar{k}_p=2.5$. The core/periphery ratio is $r=0.2$. The black dashed line in plot {\bf b} is the numerical solution of Eqs.~(\ref{eq:1}) and (\ref{eq:2}) with $\bar{k}_{cp}=0$. The inset shows the approach to the theoretical prediction at the second transition point as the size of the system is increased.}}
\label{fig:5}
\end{figure}

Figures~\ref{fig:5}~{\bf a,~b} present the simulation results of the relative size of the giant component for $\alpha=1$ and $\alpha=0.5$, respectively. In the first case, we observe a crossover effect at approximately $p=\bar{k}_p^{-1}$, whereas in the second case, we observe a clear discontinuity in the derivative of $g(p)$ at exactly $p=\bar{k}_p^{-1}$, which is consistent with the analytical prediction in Eqs.~(\ref{eq:1}) and (\ref{eq:2}) for $\bar{k}_{cp}=\bar{k}_{pc}=0$. However, the strongest evidence for the presence of a genuine double phase transition is provided by analysis of the susceptibility. In the case of a crossover effect, fluctuations in the percolated phase behave as $\langle G^2 \rangle-\langle G \rangle^2 \sim \langle G \rangle$; consequently, the quantity $\chi$ should diverge at the critical point and become size-independent after this point has been surpassed. In contrast, if the second transition in the periphery is a real phase transition, this quantity should diverge at both $p=\bar{k}_c^{-1}$ and $p=\bar{k}_p^{-1}$. This behavior is clearly observed in Figs.~\ref{fig:5}~{\bf c,~d} (we provide a finite size analysis of both transitions in the Appendix~\ref{appendix_E}).

In the case of clustered networks, it is difficult to clearly identify the core. Nevertheless, by using the giant $m1$-core as a rough approximation, we find that, in the case of $\bar{c}(k)=0.25$, the average number of connections between a node not in the giant $m1$-core and nodes in the giant $m1$-core is approximately $0.02$, indicating that the core and periphery are in fact very weakly coupled. In any case, the double divergence of $\chi$ shown in Fig.~\ref{fig:FSS}~{\bf c}, just as in the core-periphery random graph model with $\alpha<1$, is clear evidence for a genuine double phase transition.


 \section{Discussion}
As we have demonstrated, clustering has a non-trivial effect on the properties of complex networks. This effect depends on three main factors: the heterogeneity of the degree distribution, the degree-degree correlations, and the shape of the clustering spectrum $\bar{c}(k)$. If we avoid degree-degree correlations, the combination of strong clustering and heterogeneity induces the emergence of a small but macroscopic core surrounded by a large periphery. This organization redefines the percolation phase space of complex networks by inducing a new percolated phase in which the core of the network is percolated but the periphery is not. In this situation, increasing clustering makes the core larger and more entangled, thereby decreasing the percolation threshold of the first transition, as suggested in~\cite{Newman2003b,Gleeson2009,Serrano2006a}. However, in the remaining part of the network (the periphery) clustering generates small clique-like structures that are sparsely interconnected (see Fig.~\ref{fig:mcore}~c). Thus, the periphery becomes more fragile, and the percolation threshold of the second phase transition increases, in agreement with~\cite{Kiss2008,Newman2009,Miller2009,Gleeson2010a}. For weakly heterogeneous networks, the size of the core is not macroscopic; thus, clustering only makes these networks more susceptible to the removal of links. This fact reconciles the two dominant interpretations of the effect of clustering on the percolation properties of complex networks. Interestingly, this behavior is also observed in a large sample of real complex networks (see Appendix~\ref{appendix_F}), which provides evidence of the generality of this phenomenon.

We have shown that, in contrast to previous theory, it is possible to have two or more consecutive continuous phase transitions associated with the same symmetry breaking. Our work opens new lines of research concerning the effect of this core-periphery architecture on dynamical processes that occur in networks. In the case of epidemic spreading, for instance, the core could act as a reservoir of infectious agents that would be latently active in the core while the remainder of the network is uninfected.

\begin{acknowledgements}
This work was supported by a James S. McDonnell Foundation Scholar Award in Complex Systems; the ICREA Academia prize, funded by the {\it Generalitat de Catalunya}; MICINN project No.\ FIS2010-21781-C02-02; {\it Generalitat de Catalunya} grant No.\ 2014SGR608; and EC FET-Proactive Project MULTIPLEX (grant 317532).
\end{acknowledgements}

\appendix

\appendix
\section{Maximally random clustered networks}
\label{appendix_A}
Maximally random clustered networks are generated by means of a biased rewiring procedure. One edge is chosen at random that connects nodes A with B. Then, we choose at random a second link attached at least to one node (C) with the same degree of A. This link connects C with D. Then, the two edges are swapped so that nodes A and D, on the one hand, and C and B, on the other, are now connected. We take care that no self-connections or multiple connection between the same pair of nodes are created in this process. Notice that this procedure preserves both the degree of each node and the actual nodes' degrees at the end of the two original edges. Therefore, the procedure preserves the full degree-degree correlation structure encoded in the joint distribution $P(k,k')$. The procedure is ergodic and satisfies detailed balance.

Regardless of the rewiring scheme at use, the process is biased so that generated graphs belong to an exponential ensemble of graphs $\cal{G} = \mit \lbrace G \rbrace$, where each graph has a sampling probability $P(G)\propto e^{-\beta H(G)}$, where $\beta$ is the inverse of the temperature and $H(G)$ is a Hamiltonian that depends on the current network configuration. Here we consider ensembles where the Hamiltonian depends on the target clustering spectrum $\bar{c}(k)$ as
\begin{equation}
H = \sum_{k=k_{min}}^{k_c} |\bar{c}^*(k)-\bar{c}(k)|,
\end{equation}
where $\bar{c}^*(k)$ is the current degree-dependent clustering coefficient. We then use a simulated annealing algorithm based on a standard Metropolis-Hastings procedure. Let $G'$ be the new graph obtained  after one rewiring event, as defined above. The candidate network $G'$ is accepted with probability
\begin{equation}
p = \min{(1,e^{\beta [H(G)-H(G')]})} = \min{(1,e^{-\beta \Delta H})},
\end{equation}
otherwise, we keep the graph $G$ unchanged. We first start by rewiring the network $200E$ times at $\beta=0$, where $E$ is the total number of edges of the network. Then, we start an annealing procedure at $\beta_0=50$, increasing the parameter $\beta$ by a $10\%$ after $200E$ rewiring events have taken place. We keep increasing $\beta$ until the target clustering spectrum is reached within a predefined precision or no further improvement can be achieved.
\begin{figure}
\includegraphics[width=\linewidth]{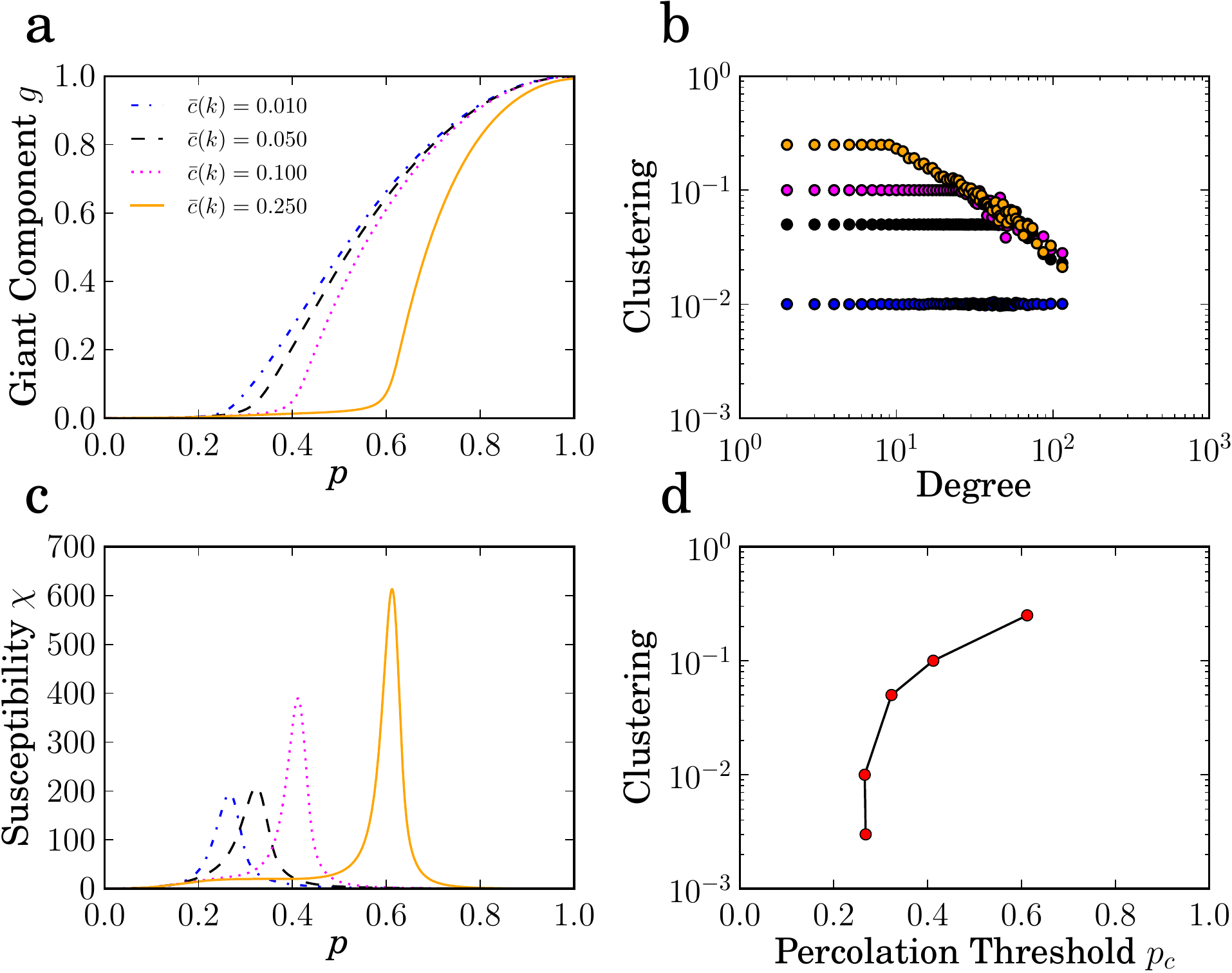}
\includegraphics[width=\linewidth]{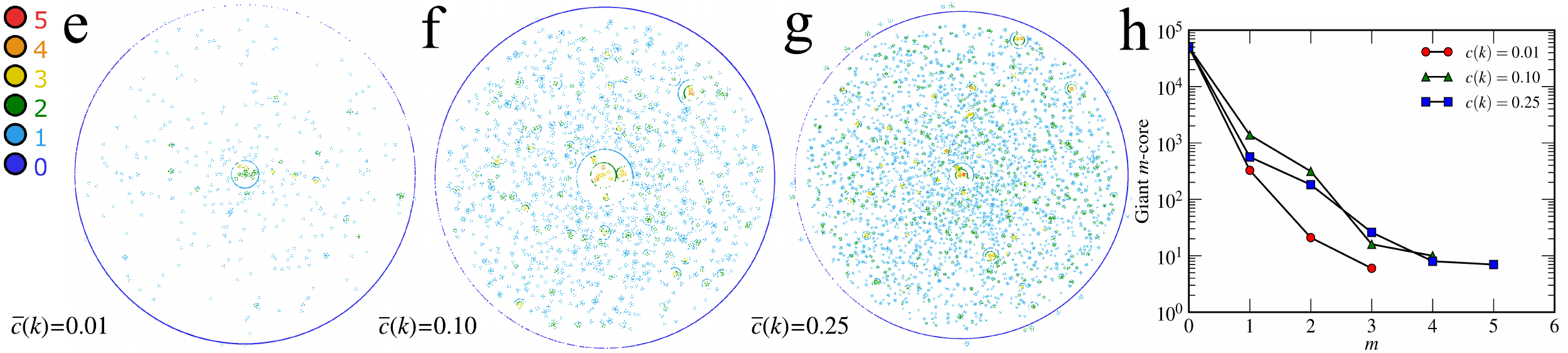}
\caption{\textbf{Top:} Bond percolation simulations for networks of $10.000$ nodes with a power law degree distribution with $\gamma=3.5$ and different levels of clustering. \textbf{a} relative size of the largest connected component $g$ as a function of the bond occupation probability $p$. \textbf{b} degree-dependent clustering coefficient $\bar{c}(k)$. \textbf{c} susceptibility $\chi$ as a function of bond occupation probability $p$. \textbf{d} Percolation threshold ($p_{max}$) as a function of the level of clustering. \textbf{Bottom: }\textbf{e-g}: $m$-core decomposition of three different networks of $50000$ nodes, $\gamma=3.5$, and different levels of clustering,  $\bar c(k) = 0.01, 0.10,0.25$. \textbf{h}: Size of the largest connected component of the m-core as a function of $m$.}
\label{fig:homo35}
\end{figure}

\begin{figure}
\includegraphics[width=\linewidth]{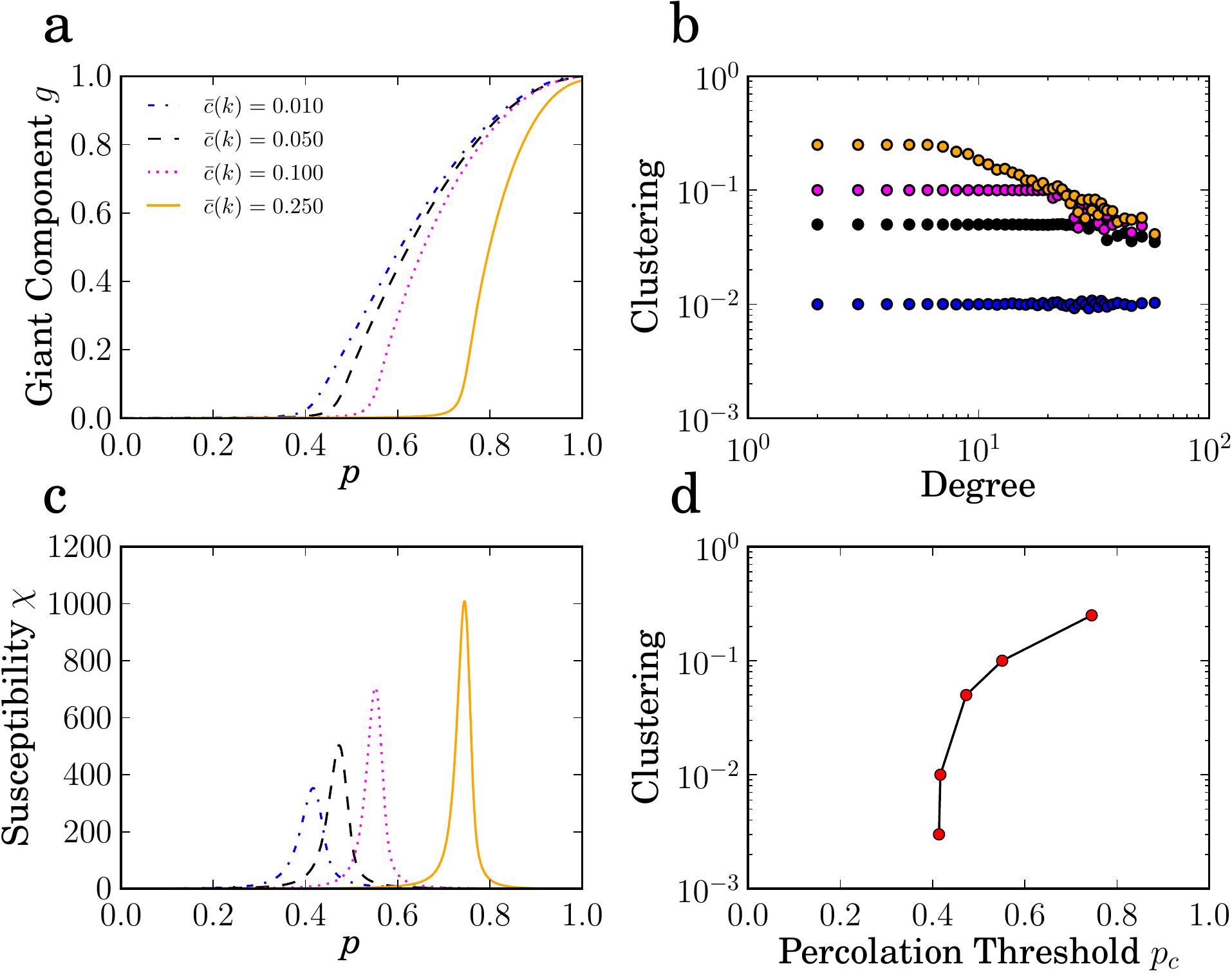}
\includegraphics[width=\linewidth]{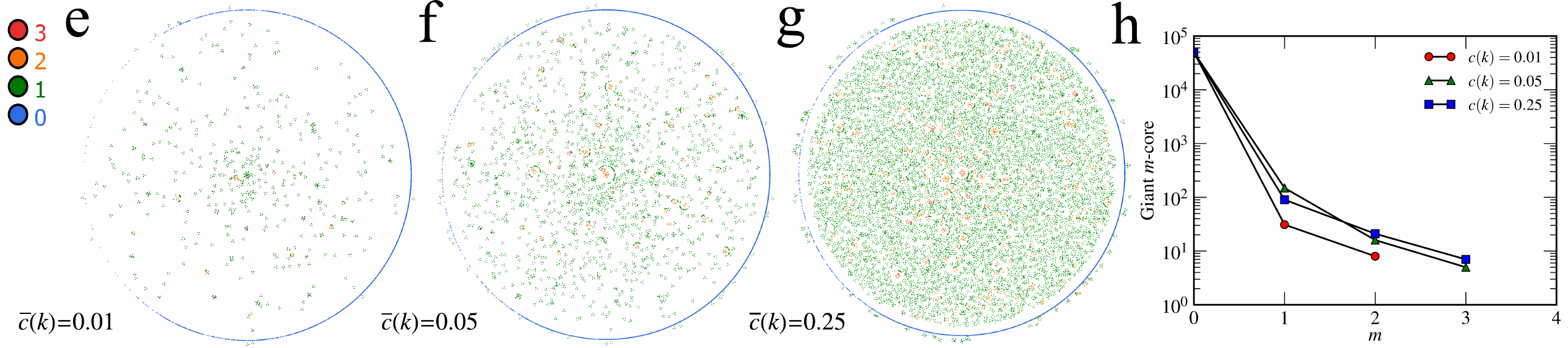}
\caption{\textbf{Top:} Bond percolation simulations for networks of $10.000$ with a power law degree distribution with $\gamma=4$ and different levels of clustering. \textbf{a} relative size of the largest connected component $g$ as a function of the bond occupation probability $p$. \textbf{b} degree-dependent clustering coefficient $\bar{c}(k)$. \textbf{c} susceptibility $\chi$ as a function of bond occupation probability $p$. \textbf{d} Percolation threshold ($p_{max}$) as a function of the level of clustering. \textbf{Bottom: }\textbf{e-g}: $m$-core decomposition of three different networks of $50000$ nodes, $\gamma=4$, and different levels of clustering,  $\bar c(k) = 0.003, 0.05,0.25$. \textbf{h}: Size of the largest connected component of the m-core as a function of $m$. }
\label{fig:homo4}
\end{figure}
\section{Effect of clustering on weakly heterogeneous networks}
\label{appendix_B}
Figures \ref{fig:homo35} for $\gamma=3.5$ and \ref{fig:homo4} for $\gamma=4$ show the comparison of the percolation properties of networks with exactly the same degree sequence and degree-degree correlations but different levels of clustering. For each network, we perform bond percolation $10^4$ times using the Newman-Ziff algorithm~\cite{Newman2000} and measure the average relative size of the largest (giant) connected component, $g \equiv \langle G \rangle/N$, and its fluctuations, {\it i.e.}, the susceptibility $\chi=[\langle G^2 \rangle - \langle G \rangle^2]/\langle G \rangle$. These results are then averaged over 50 network realizations. In finite systems, a peak in the susceptibility $\chi$ indicates the presence of a continuous phase transition and its position gives an estimate of the percolation threshold. All networks have a unique and well defined peak in $\chi$, and an increase of the clustering moves the peak to higher values of $p$. Hence clustering decreases the Giant component and increases the percolation threshold.

\section{Identification of the core}
\label{appendix_C}
\begin{figure}[t]
\includegraphics[width=\linewidth]{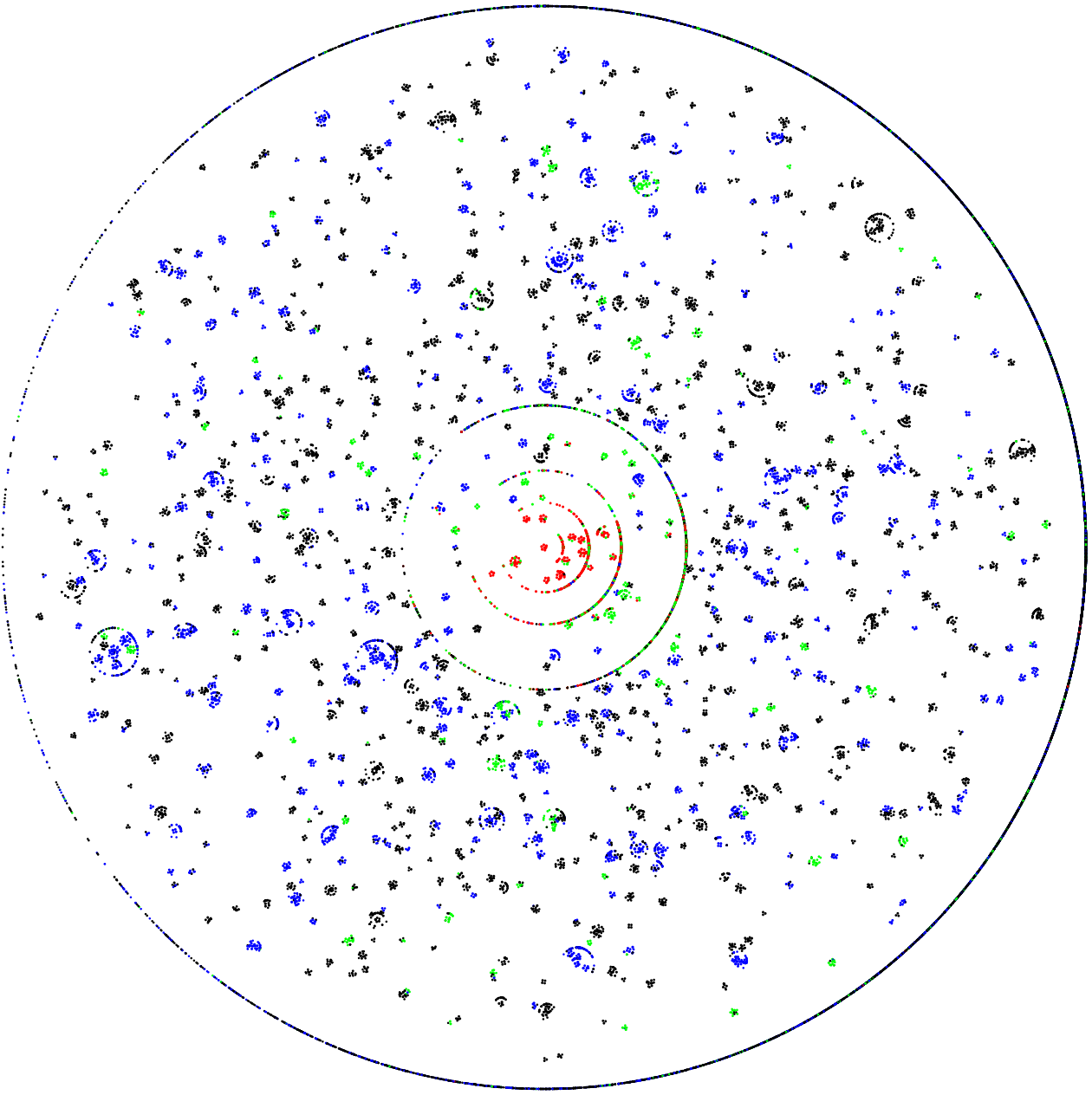}
\caption{A network of 50.000 nodes, with a power law degree distribution with $\gamma=3.1$ and a clustering spectrum $\bar c(k)=0.25$. The nodes are distributed according to its $m$-core decomposition. Red nodes (1811) are the core they because belong to the Giant component once we perform a bond percolation with $p=0.2$ (between the two percolation thresholds). Blue and green nodes are peripheral nodes that belong to the giant component at $p=0.5$ (just after the second percolation threshold). Once we subtract the core, blue nodes (10408) still remain in the GC meanwhile green nodes (4271) belong to small components. Black nodes (33510) never belong to the GC.}
\label{fig:core}
\end{figure}

In order to identify which nodes belong to the core and which to the periphery we perform a bond percolation simulation on a network of $50000$ nodes $\gamma=3.1$ and $c(k)=0.25$. We first delete all edges and then we add the edges one by one randomly. Once we added a 20\% of the total number of edges ($p=0.2$ that lays between the two percolation thresholds) the giant component (GC) defines a subgraph that we identify with the core (red nodes in Fig \ref{fig:core}). If, in the same simulation, we keep adding edges we will observe another phase transition where the periphery percolates at $p=0.5$. However the periphery has percolated regardless of the core. This can be observe if we subtract the nodes that belong to the core and see that largest component that remains is still a macroscopic component (blue nodes at Fig.~\ref{fig:core}), and only few nodes leave the GC (green nodes in Fig.~\ref{fig:core}).

\section{Bond percolation on interconnected networks}
\label{appendix_D}
Let us consider two interconnected random graphs $a$ and $b$ with average degrees $\bar{k}_{aa}$ and $\bar{k}_{bb}$, respectively. The relative size is $r=N_a/N_b$ and the average number of connections of a node in $a$ to nodes in $b$ (and vice versa) are $\bar{k}_{ab}$ and $\bar{k}_{ba}=r \bar{k}_{ab}$. 
Each node has connections to both networks and therefore its degree can be represented as a vector $\vec{k}=(k_a,k_b)$. Hence $P_a(\vec{k})$ is the probability of a node of the network $a$ to have degree $\vec{k}$ and $P_{ab}(\vec{k}'|\vec{k})$ is the probability that a node of $a$ with degree $\vec{k}$ is connected to a node of $b$ with degree $\vec{k}'$. 
The relative size of the giant component of the combined network is
\begin{equation}
g(p)=\frac{r}{1+r}g_a(p)+\frac{1}{1+r}g_b(p).
\label{eq:g(p)}
\end{equation}
Where $g_a$ is the probability that a node of $a$ belongs to the giant component, or $1$ minus the probability that it belongs to a finite cluster, that is, $g_{a} = 1 - \sum_{s=0}^{\infty}Q_{a}(s)$, where $Q_a(s)$ is the probability that a randomly chosen node from network $a$ belongs to a cluster of size $s$. 

In heterogeneous networks, the size of the cluster a given node belongs to is correlated with the degree of the node. Thus, $Q_a(s)$ must be evaluated as $Q_a(s)=\sum_{\vec{k}} P_a(\vec{k}) Q_a(s|\vec{k})$, where $Q_a(s|\vec{k})$ is the probability that a node from network $a$  of degree $\vec{k}$ belongs to a cluster of size $s$. The latter function satisfies
\begin{widetext}
\begin{equation}
\begin{split}
Q_a(s|\vec{k}) &= \sum_{n_a} {k_a \choose n_a}p^{n_a}(1-p)^{k_a-n_a}\sum_{n_b} {k_b \choose n_b}p^{n_b}(1-p)^{k_b-n_b}\\ & \sum_{s_1\cdots s_{n_a}}G_{aa}(s_1|\vec{k})\cdots G_{aa}(s_{n_a}|\vec{k})\sum_{s'_1\cdots s'_{n_b}}G_{ab}(s'_1|\vec{k})\cdots G_{ab}(s'_{n_b}|\vec{k})\\ &\delta_{s,1+s_1+\cdots +s_{n_a}+s'_1+\cdots+s'_{n_b}},
\end{split}
\end{equation}
where $G_{aa}(s|\vec{k})$ ($G_{ab}(s|\vec{k})$) is the probability to reach $s$ other nodes by following  a neighbor in network $a$ ($b$). The generating function of $Q_a(s|\vec{k})$ can be written as
\begin{equation}
\begin{split}
\hat{Q}_a(z|\vec{k}) = \sum_{s=0}^{\infty} Q_a(s|\vec{k}) z^{s} = z (1-p+p\hat{G}_{aa}(z|\vec{k}))^{k_a}(1-p+p\hat{G}_{ab}(z|\vec{k}))^{k_b} .
\end{split}
\end{equation}
Functions $G_{aa}(s|\vec{k})$, $G_{ab}(s|\vec{k})$, $G_{ba}(s|\vec{k})$, and $G_{bb}(s|\vec{k})$ follow similar recurrence equations. Thus, their generating functions satisfy
\begin{equation}
\hat{G}_{aa}(z|\vec{k}) = z \sum_{\vec{k}} P_{aa}(\vec{k}'|\vec{k}) (1-p+p\hat{G}_{aa}(z|\vec{k}))^{k_a'-1}(1-p+p\hat{G}_{ab}(z|\vec{k}))^{k_b'} 
\end{equation}
\begin{equation}
\hat{G}_{ab}(z|\vec{k}) = z \sum_{\vec{k}} P_{ab}(\vec{k}'|\vec{k}) (1-p+p\hat{G}_{ba}(z|\vec{k}))^{k_a'-1}(1-p+p\hat{G}_{bb}(z|\vec{k}))^{k_b'} 
\end{equation}
\begin{equation}
\hat{G}_{ba}(z|\vec{k}) = z \sum_{\vec{k}} P_{ba}(\vec{k}'|\vec{k}) (1-p+p\hat{G}_{aa}(z|\vec{k}))^{k_a'}(1-p+p\hat{G}_{ab}(z|\vec{k}))^{k_b'-1} 
\end{equation}
\begin{equation}
\hat{G}_{bb}(z|\vec{k}) = z \sum_{\vec{k}} P_{bb}(\vec{k}'|\vec{k}) (1-p+p\hat{G}_{ba}(z|\vec{k}))^{k_a'}(1-p+p\hat{G}_{bb}(z|\vec{k}))^{k_b'-1} ,
\end{equation}
\end{widetext}
where $P_{aa}(\vec{k}'|\vec{k})$ is the probability that a randomly chosen neighbor among all the $a$ neighbors of a node that belongs to network $a$ with degree $\vec{k}$ has degree $\vec{k}'$, and analogously for the rest of the transition probabilities. 

For networks with no degree-degree correlations, these transition probabilities simplify as  
\begin{equation}
\begin{split}
P_{aa}(\vec{k}'|\vec{k}) = \frac{k_a' P_a(\vec{k}')}{\bar{k}_{aa}} \quad
P_{bb}(\vec{k}'|\vec{k}) = \frac{k_b' P_b(\vec{k}')}{\bar{k}_{bb}} \\
P_{ab}(\vec{k}'|\vec{k}) = \frac{k_a' P_b(\vec{k}')}{\bar{k}_{ba}}  \quad
P_{ba}(\vec{k}'|\vec{k}) = \frac{k_a' P_a(\vec{k}')}{\bar{k}_{ab}}.
\end{split}
\end{equation}
This implies that functions $G_{aa}(z|\vec{k})$, $G_{ab}(z|\vec{k})$, $G_{ba}(z|\vec{k})$, and $G_{bb}(z|\vec{k})$ become independent of $\vec{k}$. We further assume that the number of neighbors from $a$ and $b$ of a given node are uncorrelated, that is
\begin{equation}
P_a(\vec{k}) = P_a(k_a)P_a(k_b)  \quad P_b(\vec{k}) = P_b(k_a)P_b(k_b).
\end{equation}
In the case of two coupled Erd\"os-R\'enyi random graphs, the degree distributions $P_a(k_a)$, $P_a(k_b)$, $P_b(k_a)$, and $P_b(k_b)$ are all Poisson distributions of parameter $\bar{k}_{aa}$, $\bar{k}_{ab}$, $\bar{k}_{ba}$, and $\bar{k}_{bb}$, respectively. In this case, it is easy to check that $\hat{Q}_a(z)=\hat{G}_{aa}(z)$,  $\hat{Q}_b(z)=\hat{G}_{bb}(z)$, and
\begin{equation}
\hat{G}_{aa}(z)= z e^{-\bar{k}_{aa} p (1 - \hat{G}_{aa}(z))}e^{-\bar{k}_{ab} p (1 - \hat{G}_{ab}(z))} 
\label{eq:Gaa}
\end{equation}
\begin{equation}
\hat{G}_{ab}(z) = z e^{-\bar{k}_{ba} p (1 - \hat{G}_{ba}(z))}e^{-\bar{k}_{bb} p (1 - \hat{G}_{bb}(z))}
\label{eq:Gab}
\end{equation}
\begin{equation}
\hat{G}_{ba}(z) = z e^{-\bar{k}_{ab} p (1 - \hat{G}_{ab}(z))}e^{-\bar{k}_{aa} p (1 - \hat{G}_{aa}(z))}
\label{eq:Gba}
\end{equation}
\begin{equation}
\hat{G}_{bb}(z) = z e^{-\bar{k}_{bb} p (1 - \hat{G}_{bb}(z))}e^{-\bar{k}_{ba} p (1 - \hat{G}_{ba}(z))}.
\label{eq:Gbb}
\end{equation}
Finally, using that $g_a=1-\hat{Q}_a(z=1)=1-\hat{G}_{aa}(z=1)$, $g_b=1-\hat{Q}_b(z=1)=1-\hat{G}_{bb}(z=1)$ and after defining $g_{ab}=1-\hat{G}_{ab}(z=1)$ and $g_{ba}=1-\hat{G}_{ba}(z=1)$ we obtain Eq.~(\ref{eq:2}).

\section{Finite size scaling of the core-periphery random graph model}
\label{appendix_E}
We first notice that the susceptibility that we use in our work is not the standard one, although it is directly related to it. The standard one is defined as
\begin{equation}
\chi_{st} \equiv \frac{\langle G^2 \rangle -\langle G \rangle^2}{N},
\end{equation}
whereas ours is defined as
\begin{equation}
\chi \equiv \frac{\langle G^2 \rangle -\langle G \rangle^2}{\langle G \rangle},
\end{equation}
For a finite system of size $N$, the peak of the susceptibility near the critical point behaves as $\chi_{st}^{max} \sim N^{\gamma/\nu}$ and the average cluster size as $\langle G \rangle \sim N^{1-\beta/\nu}$ (in this context $\gamma$ is not the exponent of the degree distribution but the critical exponent of the susceptibility). Therefore, our version of the susceptibility $\chi$ diverges near the critical point as $\chi \sim N^{\gamma'/\nu}$, where $\gamma'=\gamma+\beta$.

Let $(\beta_c,\gamma'_c,\nu_c)$ and $(\beta_p,\gamma'_p,\nu_p)$ be the critical exponents of the core and the periphery when they are isolated from each other. Close to the percolation transition of the core, the giant component is mainly composed of nodes in the core and, therefore, we expect the first transition to have the critical properties of regular percolation in the core subgraph, in particular, the susceptibility near the first peak diverges with the exponent $\gamma'_c/\nu_c$. Close to the second transition point, the giant component is the sum of the giant component in the core, $G_c$, plus the percolating cluster in the periphery, $G_p$. The susceptibility in this region can be evaluated as
\begin{equation}
\chi \approx \chi_c+\frac{\langle G_p \rangle}{\langle G_c \rangle} \chi_p.
\end{equation}
However, if the second transition point is well separated from the first one, close to this second transition $\chi_c \sim$ cte and $\langle G_c \rangle \sim N$. Then, we expect that near the second transition the susceptibility behaves as $\chi \sim N^{(\gamma'_p-\beta_p)/\nu_p}$. The critical exponents in the case of Erd\"os-Reny\'i random graphs are the mean field ones, that is, $\beta=\gamma=1$ and $\nu=3$. Therefore, in our simulations, we expect the first peak to diverge as $N^{2/3}$, the second peak as $N^{1/3}$ and the approach of the position of the peaks to their respective critical points as $p_{max} \sim p_c+A N^{-1/3}$. This is confirmed in Fig.~\ref{fig:FSS}.

\begin{figure}
\includegraphics[width=\linewidth]{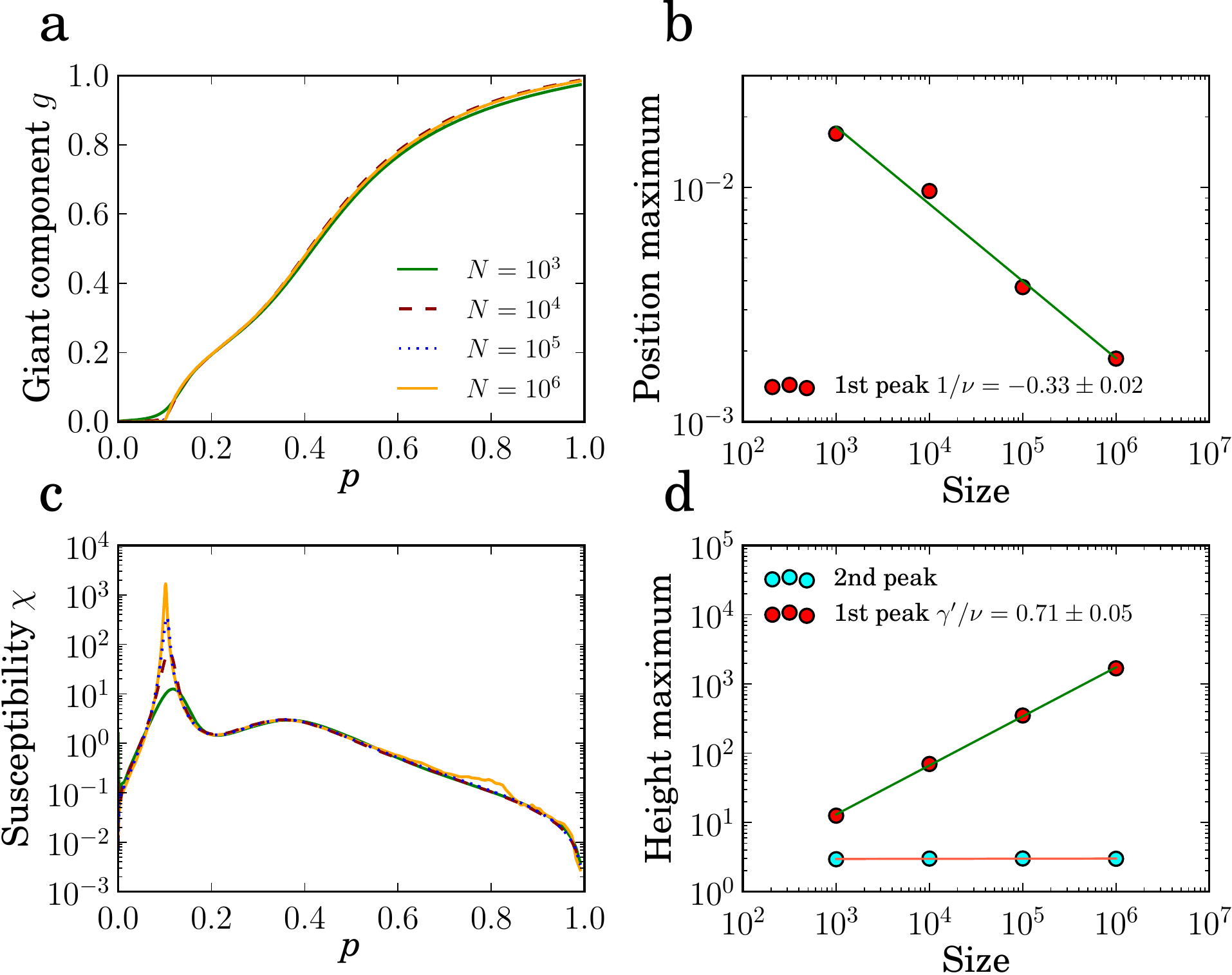}
\caption{Bond percolation simulations for the core-periphery random graph model for $\alpha=1$ for different sizes. In both cases the core has an average degree $\bar{k}_c=10$ and the periphery $\bar{k}_p=2.5$. The ratio core/periphery is $r=0.2$. {\bf a:} Relative size of the largest connected component as a function of the bond occupation probability $p$. {\bf c:} Susceptibility $\chi$ as a function of bond occupation probability $p$. {\bf b} and {\bf d:} Position $p_{max}$ and height $\chi_{max}$ of the two peaks of $\chi$ as function of network size $N$. The straight lines are power-law fits. {\bf b} and {\bf d} show the measured values of the critical exponents.}
\label{fig:FSS}
\end{figure}

\begin{figure}
\includegraphics[width=\linewidth]{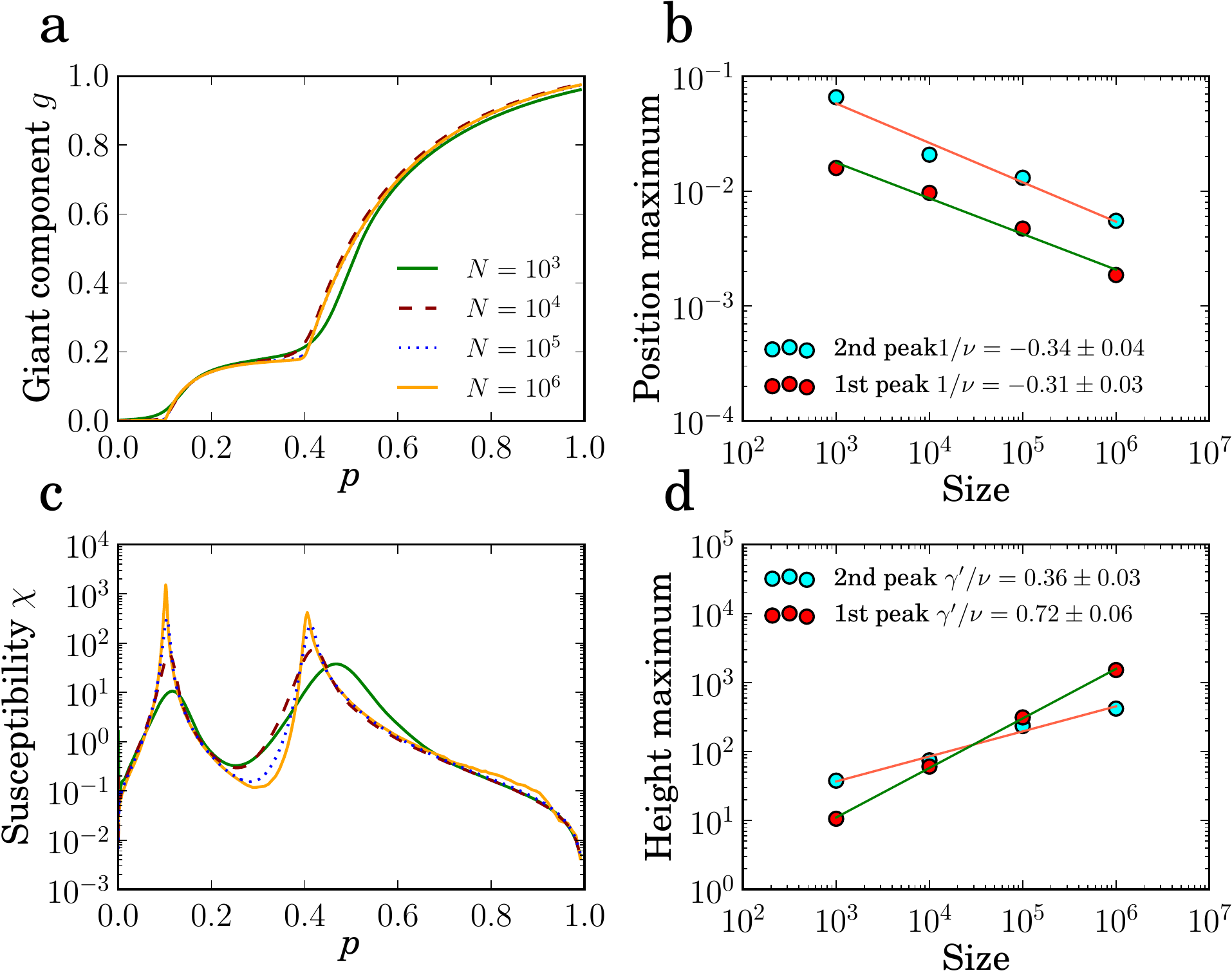}
\caption{Bond percolation simulations for the core-periphery random graph model for $\alpha=0.5$ for different sizes. In both cases the core has an average degree $\bar{k}_c=10$ and the periphery $\bar{k}_p=2.5$. The ratio core/periphery is $r=0.2$. {\bf a:} Relative size of the largest connected component as a function of the bond occupation probability $p$. {\bf c:} Susceptibility $\chi$ as a function of bond occupation probability $p$. {\bf b} and {\bf d:} Position $p_{max}$ and height $\chi_{max}$ of the two peaks of $\chi$ as function of network size $N$. The straight lines are power-law fits. {\bf b} and {\bf d} show the measured values of the critical exponents.}
\label{fig:FSS}
\end{figure}

\section{Real Networks}
\label{appendix_F}
\subsection{US air transportation network}
In the US air transportation network the nodes are airports and a link is the existence of a direct flight between two airports~\cite{Serrano2009}.
The resulting network has $583$ nodes, $1087$s edges, an average degree of $\bar k=3.73$, a clustering coefficient of $\bar C=0.43$ and a maximum degree of $k_{max}=109$.

\begin{figure*}[h]
\includegraphics[width=0.45\linewidth]{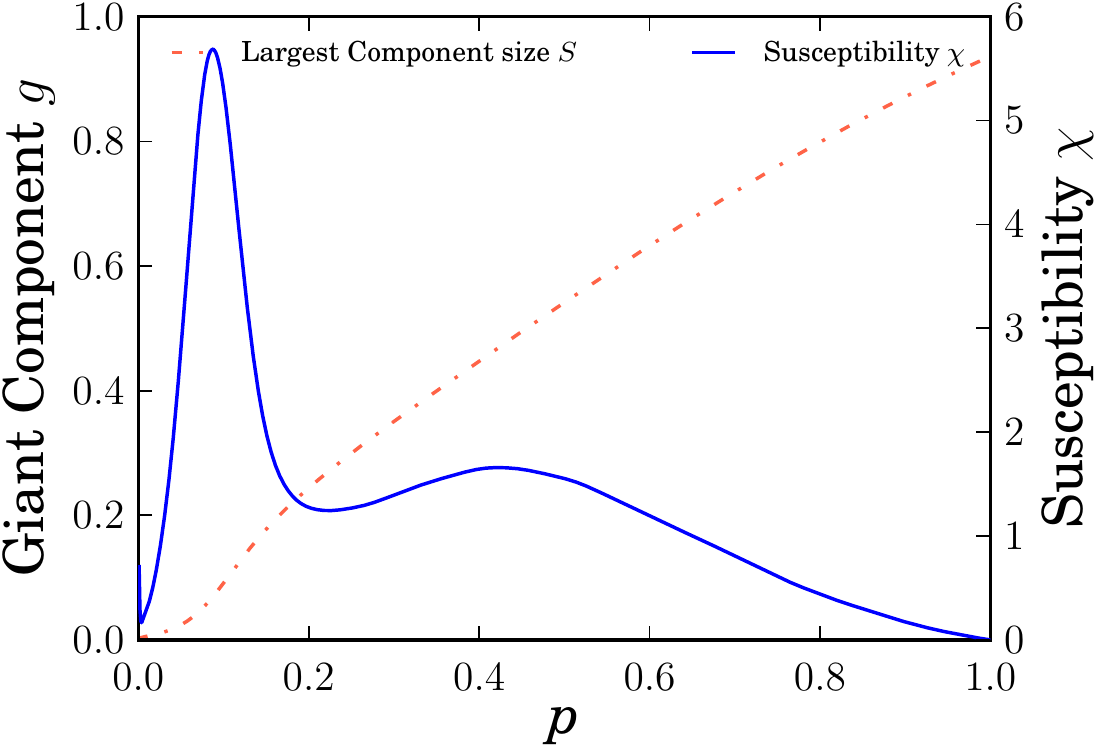}
\includegraphics[width=0.45\linewidth]{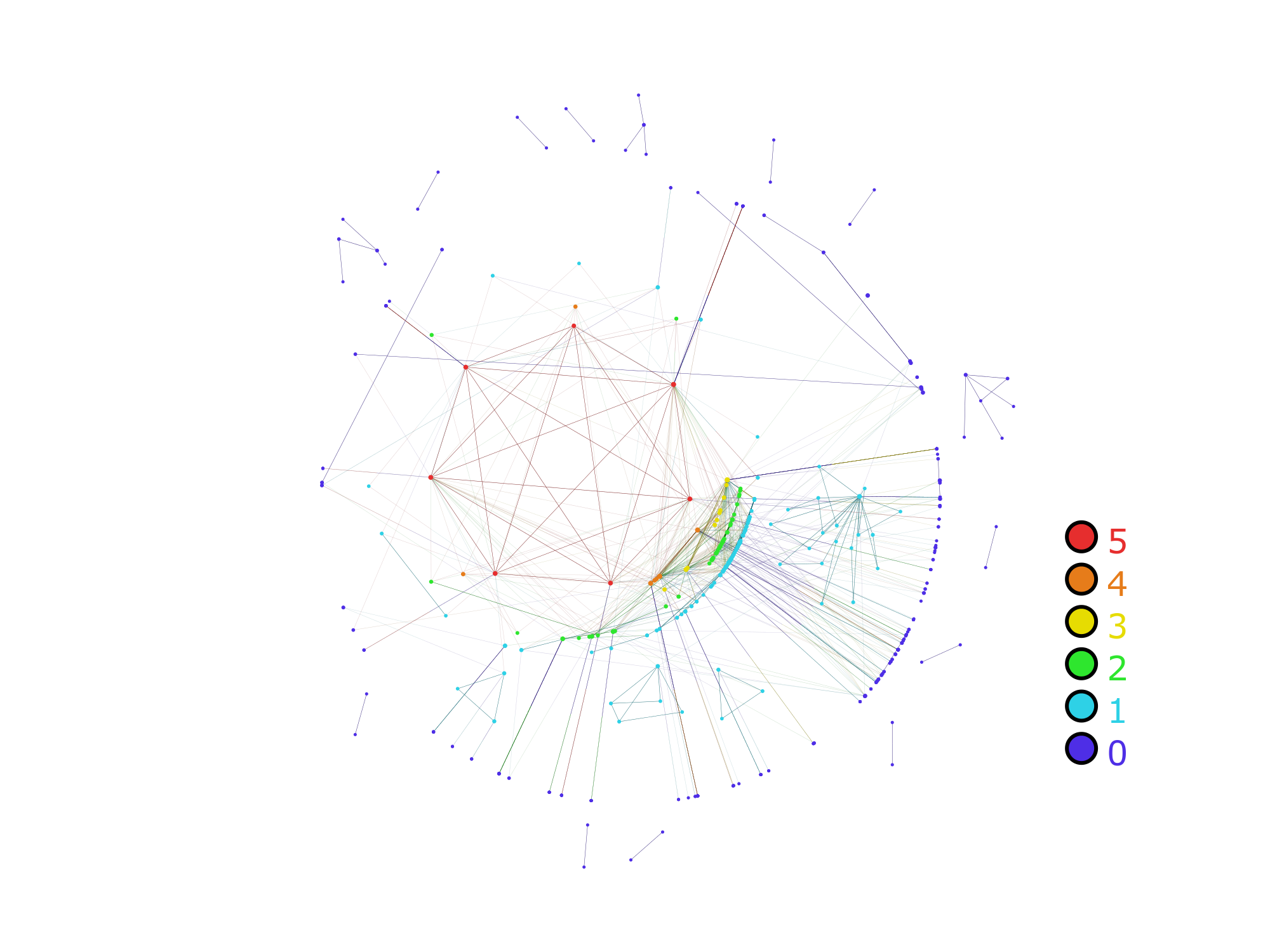}
\caption{Left: Bond percolation simulations for the US air transportation network. The relative size of the largest connected component $g$ and its susceptibility $\chi$ as a function of the bond occupation probability $p$. Right:$m$-core decomposition}
\label{fig:bond_airports}
\end{figure*}

\subsection{Human disease network}
In the "human disease network" nodes represent disorders, and two disorders are connected to each other if they share at least one gene in which mutations are associated with both disorders~\cite{Goh2007}.
The resulting network has $867$ nodes, $1527$ edges, an average degree of $\bar k=3.52$, a clustering coefficient of $\bar C=0.81$ and a maximum degree of $k_{max}=50$.
\begin{figure*}
\includegraphics[width=0.45\linewidth]{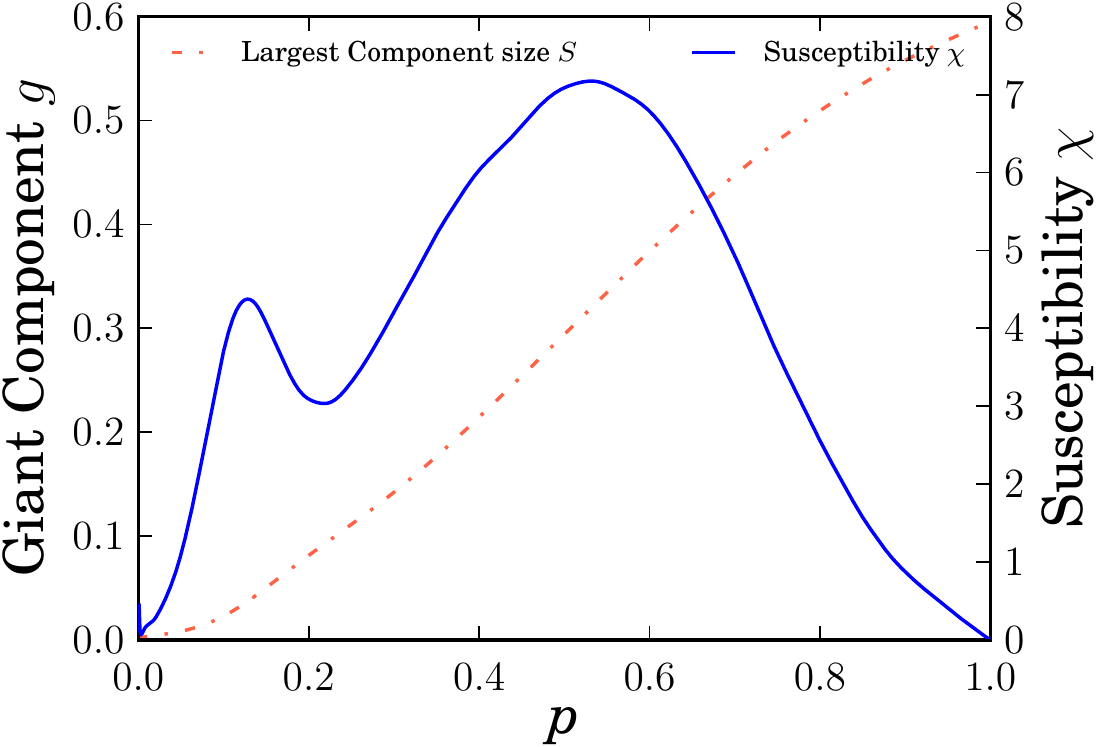}
\includegraphics[width=0.45\linewidth]{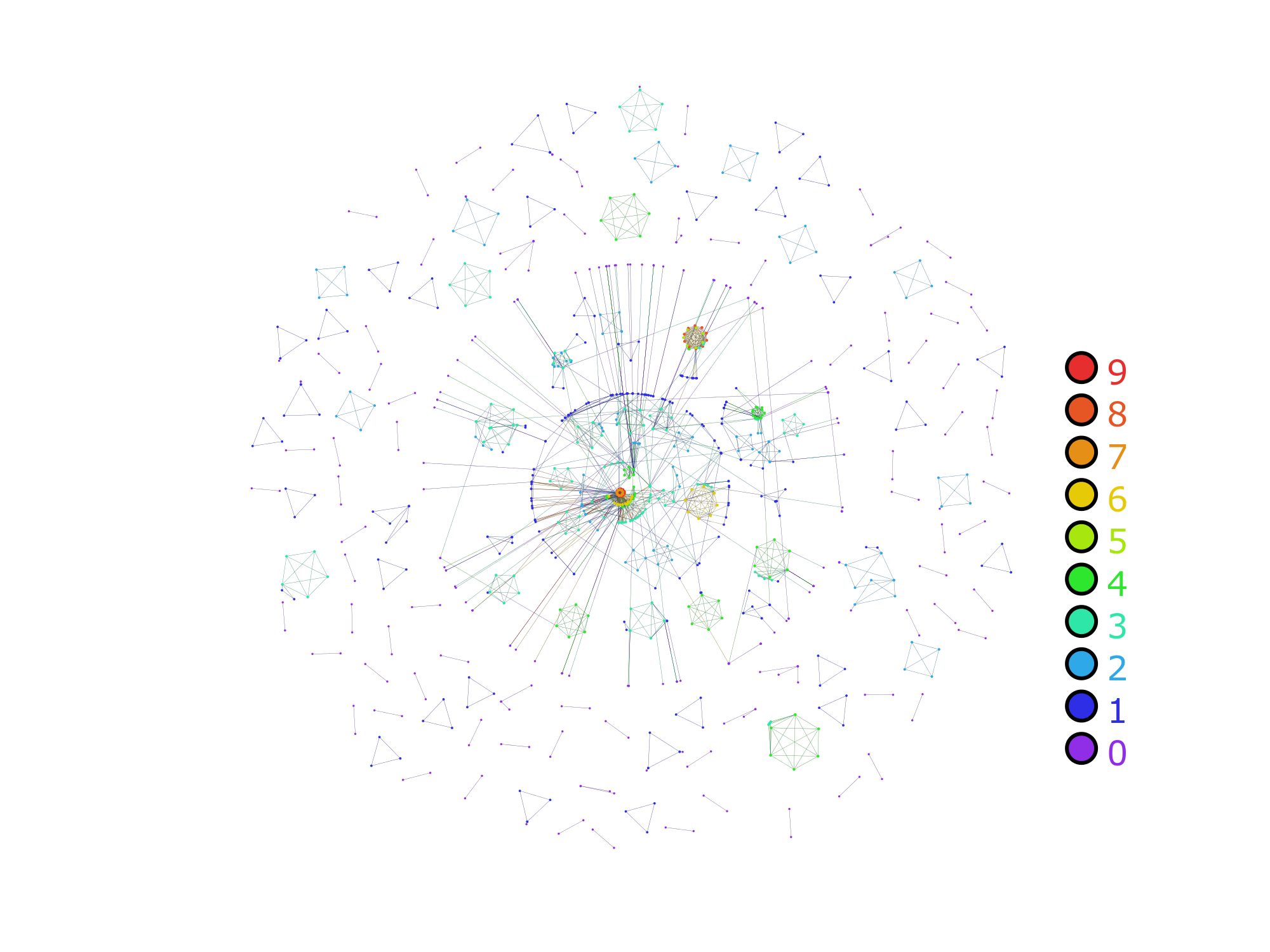}
\caption{Left: Bond percolation simulations for the human disease network network. The relative size of the largest connected component $g$ and its susceptibility $\chi$ as a function of the bond occupation probability $p$. Right:$m$-core decomposition}
\label{fig:bond_airports}
\end{figure*}

\subsection{Pokec Online Social Network}
Pokec is one of the most popular on-line social network in Slovakia. Pokec has been provided for more than 10 years and connects more than 1.6 million people by 2012. We analyse the undirected network by deleting all non-bidirectional links. For having a smaller system we only considered nodes that sign up into the online network before 2004.
The resulting network has $44285$ nodes, $75285$ edges, an average degree of $\bar k=3.4$, a clustering coefficient of $\bar C=0.09$ and a maximum degree of $k_{max}=58$.
\begin{figure*}
\includegraphics[width=0.45\linewidth]{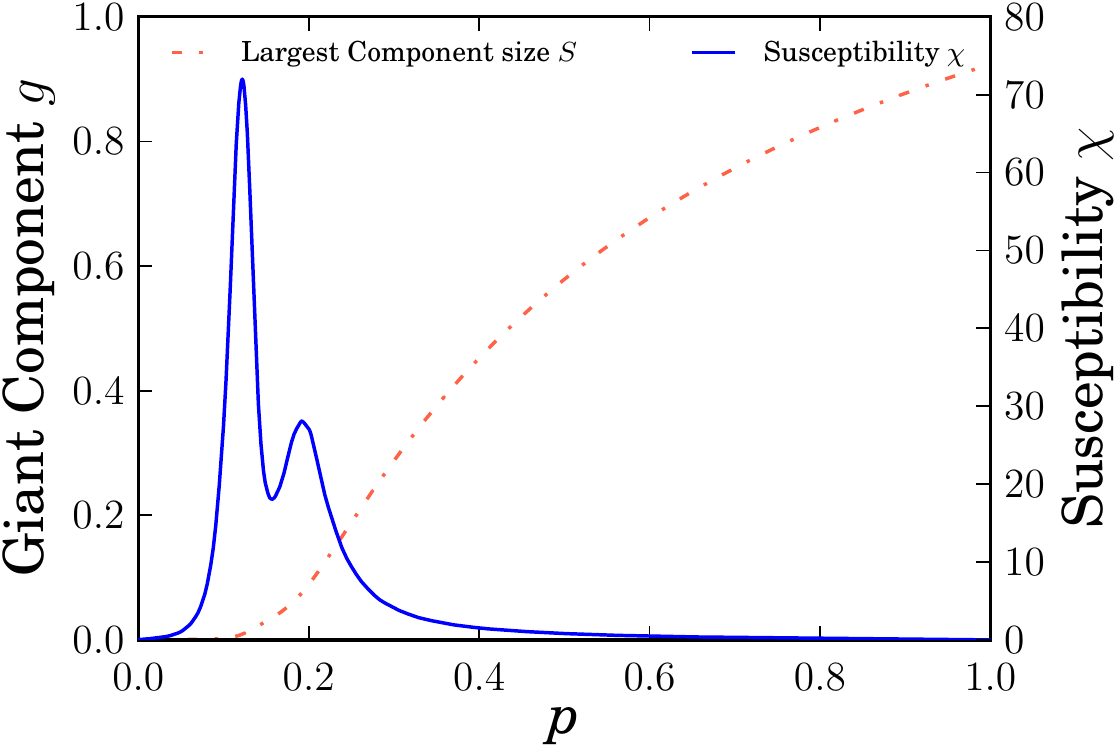}
\includegraphics[width=0.45\linewidth]{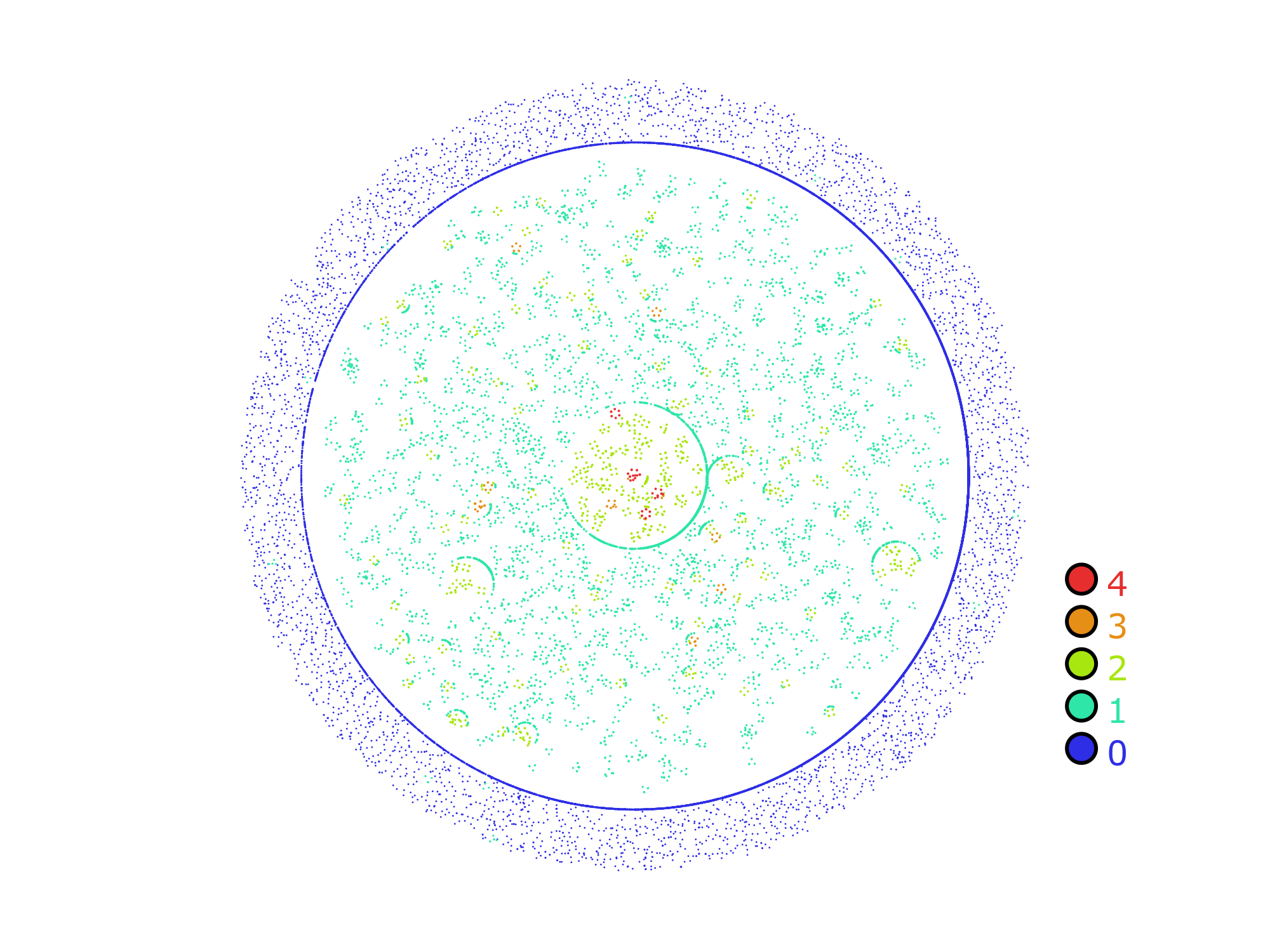}
\caption{Left: Bond percolation simulations for the Pokec Online social network. The relative size of the largest connected component $g$ and its susceptibility $\chi$ as a function of the bond occupation probability $p$. Right:$m$-core decomposition}
\label{fig:bond_airports}
\end{figure*}


\begin{thebibliography}{37}
\expandafter\ifx\csname natexlab\endcsname\relax\def\natexlab#1{#1}\fi
\expandafter\ifx\csname url\endcsname\relax
  \def\url#1{\texttt{#1}}\fi
\expandafter\ifx\csname urlprefix\endcsname\relax\def\urlprefix{URL }\fi

\bibitem[{Pastor-Satorras \& Vespignani(2001)}]{Pastor-Satorras:2001ly}
Pastor-Satorras, R. \& Vespignani, A.
\newblock Epidemic spreading in scale-free networks.
\newblock \emph{Physical Review Letters} \textbf{86}, 3200--3203 (2001).

\bibitem[{Lloyd \& May(2001)}]{Lloyd:2001mz}
Lloyd, A. \& May, R.
\newblock Epidemiology - How viruses spread among computers and people.
\newblock \emph{Science} \textbf{292}, 1316--1317 (2001).

\bibitem[{Boguna \emph{et~al.}(2003)Boguna, Pastor-Satorras \&
  Vespignani}]{Boguna:2003zr}
Boguna, M., Pastor-Satorras, R. \& Vespignani, A.
\newblock Absence of epidemic threshold in scale-free networks with degree
  correlations.
\newblock \emph{Physical Review Letters} \textbf{90} (2003).

\bibitem[{Berger \emph{et~al.}(2005)Berger, Borgs, Chayes \&
  Saberi}]{Berger:2005fk}
Berger, N., Borgs, C., Chayes, J.~T. \& Saberi, A.
\newblock On the Spread of Viruses on the Internet.
\newblock In \emph{Proceedings of the Sixteenth Annual ACM-SIAM Symposium on
  Discrete Algorithms}, SODA '05, 301--310 (Society for Industrial and Applied
  Mathematics, Philadelphia, PA, USA, 2005).


\bibitem[{Chatterjee \& Durrett(2009)}]{Chatterjee:2009uq}
Chatterjee, S. \& Durrett, R.
\newblock Contact processes on random graphs with power law degree
  distributions have critical value 0.
\newblock \emph{The Annals of Probability} \textbf{37}, 2332--2356 (2009).


\bibitem[{Bogu\~n\'a \emph{et~al.}(2013)Bogu\~n\'a, Castellano \&
  Pastor-Satorras}]{Boguna:2013kx}
Bogu\~n\'a, M., Castellano, C. \& Pastor-Satorras, R.
\newblock Nature of the Epidemic Threshold for the
  Susceptible-Infected-Susceptible Dynamics in Networks.
\newblock \emph{Phys. Rev. Lett.} \textbf{111}, 068701 (2013).

\bibitem[{Bianconi(2002)}]{Bianconi:2002fj}
Bianconi, G.
\newblock Mean field solution of the Ising model on a Barabasi-Albert network.
\newblock \emph{Physics Letters A} \textbf{303}, 166--168 (2002).

\bibitem[{Goltsev \emph{et~al.}(2003)Goltsev, Dorogovtsev \&
  Mendes}]{Goltsev:2003yq}
Goltsev, A., Dorogovtsev, S. \& Mendes, J.
\newblock Critical phenomena in networks.
\newblock \emph{Physical Review E} \textbf{67} (2003).

\bibitem[{Hinczewski \& Berker(2006)}]{Hinczewski:2006vn}
Hinczewski, M. \& Berker, A.~N.
\newblock Inverted Berezinskii-Kosterlitz-Thouless singularity and
  high-temperature algebraic order in an Ising model on a scale-free
  hierarchical-lattice small-world network.
\newblock \emph{Physical Review E} \textbf{73}, 066126 (2006).

\bibitem[{Dorogovtsev \emph{et~al.}(2008)Dorogovtsev, Goltsev \&
  Mendes}]{Dorogovtsev:2008kx}
Dorogovtsev, S.~N., Goltsev, A.~V. \& Mendes, J. F.~F.
\newblock Critical phenomena in complex networks.
\newblock \emph{Reviews of Modern Physics} \textbf{80}, 1275--1335 (2008).

\bibitem[{Cohen \emph{et~al.}(2000)Cohen, Erez, Ben-Avraham \&
  Havlin}]{Cohen2000}
Cohen, R., Erez, K., Ben-Avraham, D. \& Havlin, S.
\newblock Resilience of the Internet to random breakdowns.
\newblock \emph{Physical Review Letters} 20--22 (2000).


\bibitem[{Callaway \emph{et~al.}(2000)Callaway, Newman, Strogatz \&
  Watts}]{Callaway2000}
Callaway, D.~S., Newman, M.~E., Strogatz, S.~H. \& Watts, D.~J.
\newblock Network robustness and fragility: percolation on random graphs.
\newblock \emph{Physical Review Letters} \textbf{85}, 5468--5471 (2000).

\bibitem[{Cohen \emph{et~al.}(2002)Cohen, Ben-Avraham \& Havlin}]{Cohen2002}
Cohen, R., Ben-Avraham, D. \& Havlin, S.
\newblock Percolation critical exponents in scale-free networks.
\newblock \emph{Physical Review E} \textbf{66}, 036113 (2002).


\bibitem[{Newman(2002)}]{Newman2002}
Newman, M.
\newblock Assortative Mixing in Networks.
\newblock \emph{Physical Review Letters} \textbf{89}, 208701 (2002).
\newblock


\bibitem[{Newman \& Web(2003)}]{Newman2003}
Newman, M. E.~J. \& Web, W.-w.
\newblock Properties of highly clustered networks.
\newblock \emph{Physical Review E} 1--7 (2003).

\bibitem[{V\'{a}zquez \& Moreno(2003)}]{Vazquez2003}
V\'{a}zquez, A. \& Moreno, Y.
\newblock Resilience to damage of graphs with degree correlations.
\newblock \emph{Physical Review E} \textbf{67}, 015101 (2003).


\bibitem[{Dorogovtsev \emph{et~al.}(2001)Dorogovtsev, Mendes \&
  Samukhin}]{Dorogovtsev:2001md}
Dorogovtsev, S.~N., Mendes, J. F.~F. \& Samukhin, A.~N.
\newblock Anomalous percolation properties of growing networks.
\newblock \emph{Phys. Rev. E} \textbf{64}, 066110 (2001).

\bibitem[{Callaway \emph{et~al.}(2001)Callaway, Hopcroft, Kleinberg, Newman \&
  Strogatz}]{Callaway:2001fk}
Callaway, D.~S., Hopcroft, J.~E., Kleinberg, J.~M., Newman, M. E.~J. \&
  Strogatz, S.~H.
\newblock Are randomly grown graphs really random?
\newblock \emph{Phys. Rev. E} \textbf{64}, 041902 (2001).


\bibitem[{Buldyrev \emph{et~al.}(2010)Buldyrev, Parshani, Paul, Stanley \&
  Havlin}]{Buldyrev2010a}
Buldyrev, S.~V., Parshani, R., Paul, G., Stanley, H.~E. \& Havlin, S.
\newblock Catastrophic cascade of failures in interdependent networks.
\newblock \emph{Nature} \textbf{464}, 1025--8 (2010).


\bibitem[{Son \emph{et~al.}(2012)Son, Bizhani, Christensen, Grassberger \&
  Paczuski}]{Son:2012lq}
Son, S.-W., Bizhani, G., Christensen, C., Grassberger, P. \& Paczuski, M.
\newblock Percolation theory on interdependent networks based on epidemic
  spreading.
\newblock \emph{EPL} \textbf{97} (2012).

\bibitem[{Baxter \emph{et~al.}(2012)Baxter, Dorogovtsev, Goltsev \&
  Mendes}]{Baxter:2012fp}
Baxter, G.~J., Dorogovtsev, S.~N., Goltsev, A.~V. \& Mendes, J. F.~F.
\newblock Avalanche Collapse of Interdependent Networks.
\newblock \emph{PHYSICAL REVIEW LETTERS} \textbf{109} (2012).

\bibitem[{Kiss \& Green(2008)}]{Kiss2008}
Kiss, I. \& Green, D.
\newblock Comment on ``Properties of highly clustered networks''.
\newblock \emph{Physical Review E} \textbf{78}, 048101 (2008).


\bibitem[{Newman(2009)}]{Newman2009}
Newman, M.
\newblock Random Graphs with Clustering.
\newblock \emph{Physical Review Letters} \textbf{103}, 058701 (2009).
\newblock


\bibitem[{Miller(2009)}]{Miller2009}
Miller, J.
\newblock Percolation and epidemics in random clustered networks.
\newblock \emph{Physical Review E} \textbf{80}, 020901 (2009).


\bibitem[{Gleeson \emph{et~al.}(2010)Gleeson, Melnik \& Hackett}]{Gleeson2010a}
Gleeson, J.~P., Melnik, S. \& Hackett, A.
\newblock How clustering affects the bond percolation threshold in complex
  networks.
\newblock \emph{Physical Review E} \textbf{81}, 066114 (2010).


\bibitem[{Newman(2003)}]{Newman2003b}
Newman, M.
\newblock Properties of highly clustered networks.
\newblock \emph{Physical Review E} \textbf{68}, 026121 (2003).


\bibitem[{Gleeson(2009)}]{Gleeson2009}
Gleeson, J.
\newblock Bond percolation on a class of clustered random networks.
\newblock \emph{Physical Review E} \textbf{80}, 036107 (2009).


\bibitem[{Serrano \& Bogu\~{n}\'{a}(2006)}]{Serrano2006a}
Serrano, M. \& Bogu\~{n}\'{a}, M.
\newblock Clustering in complex networks. II. Percolation properties.
\newblock \emph{Physical Review E} \textbf{74}, 056115 (2006).


\bibitem[{Csermely \emph{et~al.}(2013)Csermely, London, Wu \&
  Uzzi}]{csermely:2013}
Csermely, P., London, A., Wu, L.-Y. \& Uzzi, B.
\newblock Structure and dynamics of core/periphery networks.
\newblock \emph{Journal of Complex Networks} \textbf{1}, 93--123 (2013).

\bibitem[{Nagler \emph{et~al.}(2012)Nagler, Tiessen \& Gutch}]{Nagler:2012fr}
Nagler, J., Tiessen, T. \& Gutch, H.~W.
\newblock Continuous Percolation with Discontinuities.
\newblock \emph{Phys. Rev. X} \textbf{2}, 031009 (2012).


\bibitem[{Chen \emph{et~al.}(2013{\natexlab{a}})}]{Chen:2013rt}
Chen, W. \emph{et~al.}
\newblock Phase transitions in supercritical explosive percolation.
\newblock \emph{Phys. Rev. E} \textbf{87}, 052130 (2013{\natexlab{a}}).


\bibitem[{Chen \emph{et~al.}(2013{\natexlab{b}})}]{Chen:2013ys}
Chen, W. \emph{et~al.}
\newblock Unstable supercritical discontinuous percolation transitions.
\newblock \emph{Phys. Rev. E} \textbf{88}, 042152 (2013{\natexlab{b}}).


\bibitem[{Bianconi \& Dorogovtsev(2014)}]{Bianconi:2014}
Bianconi, G. \& Dorogovtsev, S.~N.
\newblock Multiple percolation transitions in a configuration model of network
  of networks.
\newblock \emph{arXiv:1402.0218}  (2014).

\bibitem[{Newman \& Ziff(2000)}]{Newman2000}
Newman, M.~E. \& Ziff, R.~M.
\newblock Efficient Monte Carlo algorithm and high-precision results for
  percolation.
\newblock \emph{Physical Review Letters} \textbf{85}, 4104--7 (2000).


\bibitem[{Colomer-de Sim\'{o}n \emph{et~al.}(2013)Colomer-de Sim\'{o}n,
  Serrano, Beir\'{o}, Alvarez-Hamelin \& Bogu\~{n}\'{a}}]{Colomer-de-Simon2013}
Colomer-de Sim\'{o}n, P., Serrano, M.~A., Beir\'{o}, M.~G., Alvarez-Hamelin,
  J.~I. \& Bogu\~{n}\'{a}, M.
\newblock Deciphering the global organization of clustering in real complex
  networks.
\newblock \emph{Scientific reports} \textbf{3}, 2517 (2013).


\bibitem[{Melnik \emph{et~al.}(2011)Melnik, Hackett, Porter, Mucha \&
  Gleeson}]{Melnik2011}
Melnik, S., Hackett, A., Porter, M.~a., Mucha, P.~J. \& Gleeson, J.~P.
\newblock The unreasonable effectiveness of tree-based theory for networks with
  clustering.
\newblock \emph{Physical Review E} \textbf{83}, 036112 (2011).


\bibitem[{Serrano \& Bogu{\~{n}}{\'{a}}(2006)}]{Serrano:2006qj}
Serrano, M.~A. \& Bogu{\~{n}}{\'{a}}, M.
\newblock Clustering in complex networks. I. General formalism.
\newblock \emph{Phys. Rev. E} \textbf{74}, 056114 (2006).

\bibitem{Newman2000}
\bibinfo{author}{Newman, M.~E.} \& \bibinfo{author}{Ziff, R.~M.}
\newblock \bibinfo{title}{{Efficient Monte Carlo algorithm and high-precision
  results for percolation.}}
\newblock \emph{\bibinfo{journal}{Physical review letters}}
  \textbf{\bibinfo{volume}{85}}, \bibinfo{pages}{4104--7}
  (\bibinfo{year}{2000}).


\bibitem{Serrano2009}
\bibinfo{author}{Serrano, M.~A.}, \bibinfo{author}{Bogu\~{n}\'{a}, M.} \&
  \bibinfo{author}{Vespignani, A.}
\newblock \bibinfo{title}{{Extracting the multiscale backbone of complex
  weighted networks.}}
\newblock \emph{\bibinfo{journal}{Proceedings of the National Academy of
  Sciences of the United States of America}} \textbf{\bibinfo{volume}{106}},
  \bibinfo{pages}{6483--8} (\bibinfo{year}{2009}).


\bibitem{Goh2007}
\bibinfo{author}{Goh, K.-I.} \emph{et~al.}
\newblock \bibinfo{title}{{The human disease network.}}
\newblock \emph{\bibinfo{journal}{Proceedings of the National Academy of
  Sciences of the United States of America}} \textbf{\bibinfo{volume}{104}},
  \bibinfo{pages}{8685--90} (\bibinfo{year}{2007}).


\end{thebibliography}
\end{document}